\documentclass[runningheads]{llncs}

\AtBeginDocument{%
  \providecommand\BibTeX{{%
    \normalfont B\kern-0.5em{\scshape i\kern-0.25em b}\kern-0.8em\TeX}}}

\usepackage{times}
\usepackage{soul}
\usepackage{url}
\usepackage{amsmath}
\usepackage{booktabs}
\usepackage{algorithm}
\usepackage{algpseudocode}

\usepackage{multirow}
\usepackage{caption}
\usepackage{graphicx}
\usepackage{booktabs}
\usepackage{nicefrac}

\usepackage{textcomp}

\usepackage{xcolor}

\usepackage[colorlinks,allcolors=blue]{hyperref}
\usepackage{orcidlink}
\usepackage{comment}

\usepackage{amsmath}
\usepackage{mathtools}
\usepackage{multicol}

\usepackage{xfp}

\newcommand{\etal}{\textit{et al. }}

\usepackage[caption=false]{subfig}
\begin{document}

\title{Robust Brain Age Estimation via Regression Models and MRI-derived Features}

\author{Mansoor Ahmed\inst{1}\orcidlink{0000-0002-3614-4124} \and
Usama Sardar\inst{1}\orcidlink{0000-0002-9274-3797} \and
Sarwan Ali\inst{2} \orcidlink{0000-0001-8121-2168}\and
Shafiq Alam\inst{3} \orcidlink{0000-0002-9566-8040}\and
Murray Patterson\inst{2,*}\orcidlink{0000-0002-4329-0234} \and
Imdad Ullah Khan\inst{1,*} \orcidlink{0000-0002-6955-6168}}
\authorrunning{\textit{M. Ahmed, U. Sardar, S. Ali et al.}}

\institute{Lahore University of Management Sciences, Lahore, Pakistan\\
\email{\{mansoorbaloch931,usamasardar2022\}@gmail.com, imdad.khan@lums.edu.pk} \and
Georgia State University, Atlanta, GA, USA \\
\email{sali85@student.gsu.edu, mpatterson30@gsu.edu} \and
Massey University, Auckland, New Zealand\\
\email{salam1@massey.ac.nz} * Corresponding authors, Equal contribution}

\maketitle 
\begin{abstract}
The determination of biological brain age is a crucial biomarker in the assessment of neurological disorders and understanding of the morphological changes that occur during aging. Various machine learning models have been proposed for estimating brain age through Magnetic Resonance Imaging (MRI) of healthy controls. However, developing a robust brain age estimation (BAE) framework has been challenging due to the selection of appropriate MRI-derived features and the high cost of MRI acquisition. In this study, we present a novel BAE framework using the Open Big Healthy Brain (OpenBHB) dataset, which is a new multi-site and publicly available benchmark dataset that includes region-wise feature metrics derived from T1-weighted (T1-w) brain MRI scans of $3965$ healthy controls aged between 6 to 86 years. Our approach integrates three different MRI-derived region-wise features and different regression models, resulting in a highly accurate brain age estimation with a Mean Absolute Error (MAE) of $3.25$ years, demonstrating the framework's robustness. We also analyze our model's regression-based performance on gender-wise (male and female) healthy test groups.
The proposed BAE framework provides a new approach for estimating brain age, which has important implications for the understanding of neurological disorders and age-related brain changes.

\end{abstract}

\section{Introduction}
The biological brain age, or simply brain age, is the estimated brain age (in years) that quantifies how old a person's brain is. In contrast, the chronological age is the person's actual age relative to the calendar birth date. We call the difference between the brain age and the chronological age as Brain Estimated Age Difference (Brain-EAD). The brain age of people with various neurological disorders is known to be different from its chronological brain age and is also an important biomarker for neurodegenerative disorders~\cite{Cole2017_BA,Gaser2013_BA}. For instance, Brain-EAD of Alzheimer's Disease (AD) patients is greater than Parkinson's Disease (PD) patients~\cite{Beheshti_618_BA}. 

Brain-EAD is a data-driven biomarker that exploits brain features derived from neuroimaging modalities such as brain MRIs by leveraging machine learning algorithms for brain age prediction~\cite{franke_BA_2019}. A major challenge in developing efficient BAE frameworks is the selection of appropriate features that fully capture healthy aging and, ultimately, the machine learning model~\cite{Ediri_2020_vbm}. Existing BAE approaches have widely used region-wise features such as the global brain volumes~\cite{Farokhian_2017_aging,Cole2017_BA_CNN} or cortical measurements \cite{Aycheh_2018_bae_region,Baecker_2021_bae_reg,BEHESHTI_BA_2022,Basodi2022_BA,LIU_2022_BA} such as cortical thickness, volume, surface area, and curvatures yielding lesser accuracy. Some other approaches have used very high-dimensional voxel-wise features requiring computationally expensive resources for dimensionality reduction, such as Principal Component Analysis \cite{Baecker_2021_bae_reg}. To address these limitations, we propose the fusion of the region-wise features, i.e., global Gray Matter (GM), White Matter (WM), and Cerebrospinal Fluid (CSF) volumes and cortical measurements of parcellated regions of interest (ROI), for building a BAE framework. Therefore, we hypothesize that integrating the region-wise features may carry potentially complementary information about brain age, resulting in an improved BAE model.

In this paper, we build a BAE framework using three different sets of region-wise features derived from the T1-weighted MRI of healthy individuals by training a Generalized Linear regression Model (GLM). We evaluated the model performance on the three separate feature sets and compared them with the integrated feature set. Additionally, we compare the model performance on gender-wise (male and female) healthy test groups. Our experiments demonstrate that our model outperforms previous methods of efficiently estimating brain age from T1-weighted MRI scans.
\par The significant contributions of this paper are the following:
\begin{enumerate}
    \item We propose the first brain age estimation model that integrates the distinct region-wise features, i.e., global GM and CSF volumes and cortical measurements of Desikan~\cite{Desikan_2006} and Destrieux~\cite{Fischl_2004} ROI.
    \item We compare the performance of BAE frameworks developed using individual region-wise metrics derived from T1-w MRI and on the gender-wise (male/female) hold-out test sets.
    \item Our model achieves improved BAE accuracy compared to the previously known BAE models using region-wise features derived from T1-w brain MRIs.
\end{enumerate}

The rest of the paper is organized as follows: Section~\ref{relatedwork} provides a review of the BAE framework. Section~\ref{model} provides the details of our proposed models. Section~\ref{experiments} details the experimental design, and Section~\ref{results} analyzes and discusses the experimental results and comparisons of different models. Finally, Section~\ref{sec:conclusion} concludes the paper.

\section{Related Work} \label{relatedwork}
In this section, we introduce the brain age estimation (BAE) problem, a comparison of features derived from MRI for BAE, multi-site MRI studies, and evaluation metrics used for assessing the performance of BAE models. 

\subsection{BAE Framework}
Biological brain age is determined by training a supervised machine learning model on MRIs of cognitively healthy subjects. This model takes input features (extracted from MRIs) as independent variables and outputs the brain age (in years) as the dependent variable. Since all the subjects are ``healthy", their Brain-EAD is assumed to be zero. Thus, the BAE model attempts to fit a function (of features) for age. A subject's brain age difference, Brain-EAD, or brain age gap, is the difference between the estimated biological and chronological ages. If the observed Brain-EAD is (significantly) higher than $0$, then the subject may have a certain underlying neurological disorder (aging their brain faster)~\cite{Beheshti_618_BA}. In contrast, it has been observed that the Brain-EAD of long-term meditation practitioners is (significantly) less than $0$ (slower rate of brain aging)~\cite{Luders_2016_BA}. 

Analysis of T1-weighted MRIs is commonly employed as they provide high-resolution images of the GM and WM~\cite{Mikheev_2008_T1-w} while providing information about atrophy in the brain's anatomical structures~\cite{fneur_2020}. Some studies also used multi-modal MRIs by integrating structural and functional MRI (fMRI) to estimate brain age~\cite{Basodi2022_BA,TAYLOR_2022_BAE}. To build BAE models, different machine learning algorithms such as Support Vector Regression (SVR)~\cite{Basodi2022_BA,BEHESHTI_BA_2022}, Gaussian Process Regression (GPR)~\cite{Baecker_2021_bae_reg,Aycheh_2018_bae_region}, Relevance Vector Regression (RVR)~\cite{FRANKE_2010_BA}, and Generalized Linear Models (GLM)~\cite{Modabbernia_2022_bae_ml} have been applied to the features extracted from T1-w MRI~\cite{LEE_2021_BA}. Similarly, researchers have applied deep learning algorithms, such as 3D convolutional neural networks (CNNs), to predict brain age using voxel-wise features extracted from GM and WM segmented T1-weighted images~\cite{Cole2017_BA,Jonsson_2019_BA}. 

\subsection{Features Selection}
MRIs contain valuable structural and functional information about alterations in the human brain, such as the reduction in global volumes with age. It has been observed that GM volume monotonically decreases from the 20s to the 70s, WM volume shows minor changes, while CSF volumes, conversely, increase from the 20s to the 70s~\cite{Taki_aging_2011}. Capturing the morphological similarities and alterations in the individual brain with age is necessary for building accurate BAE models and varies greatly with the structural measures being used~\cite{Ediri_2020_vbm}. Currently, different local and global brain features are derived from the T1-weighted MRI for brain age estimation. These features are broadly categorized into region-wise~\cite{Lee_2022_bae,LIU_2022_BA}, voxel-wise~\cite{FRANKE_2010_BA,NENADIC_2017_bae}, and surface-based metrics~\cite{Sajedi_2019_bae_survey}.  
\par The model's accuracy was degraded when individual features derived from T1-w MRI were used. Franke \etal predicted brain age using the voxel-wise features from T1-w MRI with an \textsc{mae} of $4.98$ years~\cite{FRANKE_2010_BA}. Similarly, Cole \etal used GM and WM volumes and demonstrated the \textsc{mae} to be $4.65$ years~\cite{Cole2017_BA_CNN}. Liu \etal built a BAE framework using region-wise cortical measurements and attained a Mean Absolute Error (\textsc{mae}) of $3.73$ years~\cite{LIU_2022_BA}. On the contrary, promising results were obtained when these features were integrated \cite{Baecker_2021_bae_reg,Cole2017_BA}. Authors in~\cite{BEHESHTI_BA_2022} and~\cite{Baecker_2021_bae_reg} integrated gray matter voxel-wise maps and region-wise metrics and achieved \textsc{mae} of $4.63$ and $3.7$ years respectively. The authors of~\cite{Jonsson_2019_BA} combined surface-based and voxel-wise features and showed that prediction accuracy decreases compared to individual features. Still, there is no clear understanding of the selection of appropriate features, and it can be assessed that integrating the different region-wise features derived from the T1-w brain MRIs results in improved BAE models.

\section{Proposed Model} \label{model}

In this section, we first analyze and describe the different MRI-derived region-wise features and classify participants into different age groups or clusters. Then, we build our BAE model by integrating the three region-wise features and evaluate the performance on an independent healthy test set. 
\subsection{Participants and Exclusion Criteria}
The OpenBHB dataset~\footnote{The dataset was provided (in part) by Neurospin, CEA, France.} contains the MRI-derived features of a total of $3983$ HCs collected from $10$ different sources with \> $60$ individual MRI acquisition sites. These data sources are Autism Brain Imaging Data Exchange (ABIDE-1 and ABIDE-2)~\footnote{\url{http://fcon_1000.projects.nitrc.org/indi/abide/}}, brain Genomics Superstruct Project (GSP)~\footnote{\url{https://www.nitrc.org/projects/gspdata}}, consortium for Reliability and Reproducibility (CoRR)~\footnote{\url{http://fcon_1000.projects.nitrc.org/indi/CoRR/html/}}, information eXtraction from Images (IXI)~\footnote{\url{http://brain-development.org/ixi-dataset/}}, brainomics/Localizer~\footnote{\url{https://osf.io/vhtf6/wiki/Localizer/}}, MPI-Leipzig~\footnote{\url{https://openneuro.org/datasets/ds000221/}}, narratives (NAR)~\footnote{\url{https://openneuro.org/datasets/ds002345/}}, neuroimaging Predictors of Creativity in Healthy Adults (NPC)~\footnote{\url{https://openneuro.org/datasets/ds002330/}}, and the reading Brain Project L1 Adults (RBP)~\footnote{\url{https://openneuro.org/datasets/ds003974/}}.
We excluded $18$ samples from our analysis because of duplicate IDs with conflicting ages and used the remaining dataset ($m = 3965$). The dataset has an overall uniform gender distribution with an equal distribution in each of the $10$ age bins (male = $2079$, female = $1886$) (see Figure~\ref{fig:age_distribution}).  

\begin{table}[h!]
\begin{minipage}[b]{0.48\textwidth}
	\centering
  \includegraphics[width=0.92\textwidth]{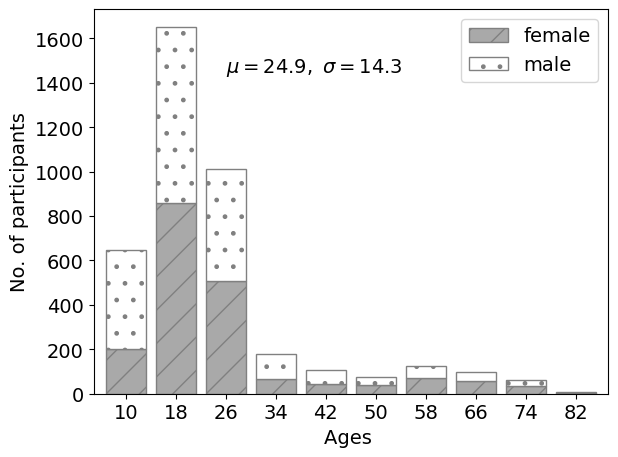}
    \captionof{figure}{Age and gender distribution of the participants.}
    \label{fig:age_distribution}
\end{minipage}
\hfill
\begin{minipage}[b]{.48\textwidth}
 \centering
 \footnotesize
  \resizebox{0.98\textwidth}{!}{
  \begin{tabular}{lcc}
    \toprule
    & Training set &Test set\\
    \midrule \midrule
    No. of HC & 3172 & 793\\
    Age $\pm$ std (years) & 25.2 $\pm$ 14.6 &23.8 $\pm$ 12.8 \\
    Sex (M$/$F) &1661$/$1548 & 434$/$359 \\
  \bottomrule
  \medskip
\end{tabular}}
\caption{Demographic characteristics of the training and test set subjects.}
  \label{tab:freq}
\end{minipage}
\end{table}

\subsection{Sampling Method}
An $80-20\%$ random split was followed to divide the data into training and testing sets with uniform gender distribution in both sets. The test set was further divided into male and female hold-out sets, as shown in Table~\ref{tab:freq}.
 
\subsection{MRI Processing}
The MRIs of all HC  are uniformly pre-processed using FreeSurfer and CAT12, while a semi-automatic quality control was also performed on the images before extracting different features~\cite{DUFUMIER_OpenBHB_2022}. Later, the volumetric measurements of different brain atlases or regions of interest (ROI) were computed.
\subsubsection{CAT12 ROI}
\begin{figure}[h!]
    \centering
    \includegraphics[scale=0.4]{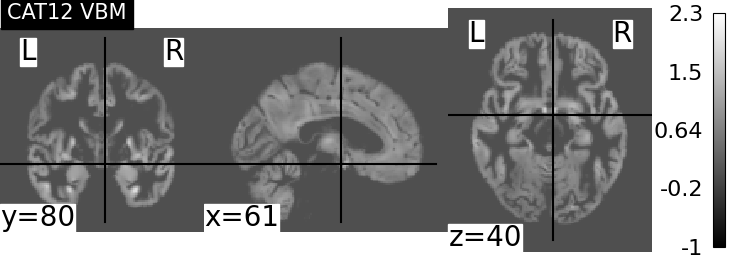}
    \caption{An illustration of GM maps generated by CAT12.}
    \label{fig:Neuromorphomterics}
\end{figure}

The CAT12 Voxel-Based Morphometry (VBM) pipeline, as explained in~\cite{Gaser_2016_cat12}, was followed, which includes non-linear registration to $1.5$ $mm^3$ MNI template, Gray Matter (GM), White Matter (WM), and Cerebrospinal Fluid (CSF) tissue segmentation, bias correction of intensity non-uniformities, and segmentation modulation by scaling with the number of volume changes due to spatial registration~\cite{DUFUMIER_OpenBHB_2022}. We used GM and CSF volumes averaged on the Neuromorphometrics atlas comprising the volumes of $142$ cortical and sub-cortical regions of both hemispheres (see Figure~\ref{fig:Neuromorphomterics}).

\subsubsection{Desikan ROI}
Using FreeSurfer, the MRIs are pre-processed by intensity normalization, skull stripping, segmentation of GM and WM, hemispheric-based tessellations, topology corrections, inflation, and registration to the ``fsaverage" template~\cite{DUFUMIER_OpenBHB_2022}. Then, the measurements for the widely-used cortical parcellations Desikan-Killiany atlas~\cite{Desikan_2006} (automated labeling of the human cerebral cortex into the gyral-based region of interest) containing the surface area ($mm^2$), GM volume ($mm^3$), cortical thickness ($mm$), thickness std dev ($mm$), 
integrated rectified mean curvature ($mm^{-1}$),
integrated rectified Gaussian curvature ($mm^{-2}$), and the intrinsic curvature index of each brain region are obtained. The two principal curvatures of each regional surface are computed and represented as $k_1$ and $k_2$. Hence, 
\begin{gather*}
    \text{Mean curvature} = \nicefrac{1}{2}(k_1 + k_2) \qquad
    \text{Gaussian curvature} = k_1 \times k_2     
\end{gather*}
  These measurements for Desikan parcellation ($2 \times 7 \times 34$) were used to train the BAE model. We call these features set ``Desikan-ROI''. 
\subsubsection{Destrieux ROI}
The MRI preprocessing pipeline is the same as Desikan ROI, with varying brain regions in both brain parcellations. The features constitute the seven cortical measurements (as explained in Desikan ROI) of the Destrieux atlas~\cite{Fischl_2004} comprising $74$ global regions of interest for each hemisphere.

We call the concatenation of Desikan ROI, Destrieux ROI, and CAT12 ROI ``all region-wise'' metrics.

\subsection{Feature Engineering and Model Selection}

Let \(X^{m \times n} \in \mathbf{R}^{m \times n}\) be the data matrix of HCs with \(m\) being the number of subjects and \(n\) being the MRI-derived features while \(Y^{m \times 1}\) be their chronological ages. 

We visualized the feature separation for the three age groups (adults, adolescents, and elders) computed through $k$-means clustering using $t$-distributed Stochastic Neighbor Embedding ($t$-SNE)~\cite{van2008visualizing} plots. The summary of $k$-means clustering of ages is presented in Table~\ref{clusters}.

\begin{table}[h!]
\centering
     \footnotesize
    \begin{tabular}{cccccc}
         \toprule
         Cluster ID&Min &Max &Mean &Frequency &Group\\ \midrule \midrule
             0 & 6 & 17 & 11.6 & 620 &Adolescents\\
             2 & 17 & 42 & 23 & 2223 &Adults\\
              1 & 42 & 86 & 61.1 & 366 &Elders\\
            \bottomrule
        \end{tabular} 
        \medskip
        \caption{Summary of 3-means clustering of the participants' ages (in years).}
        \label{clusters}
\end{table}

\begin{figure}[h!]
     \centering    
    \begin{minipage}{0.48\textwidth}
        \centering
        \includegraphics[width=.88\linewidth]{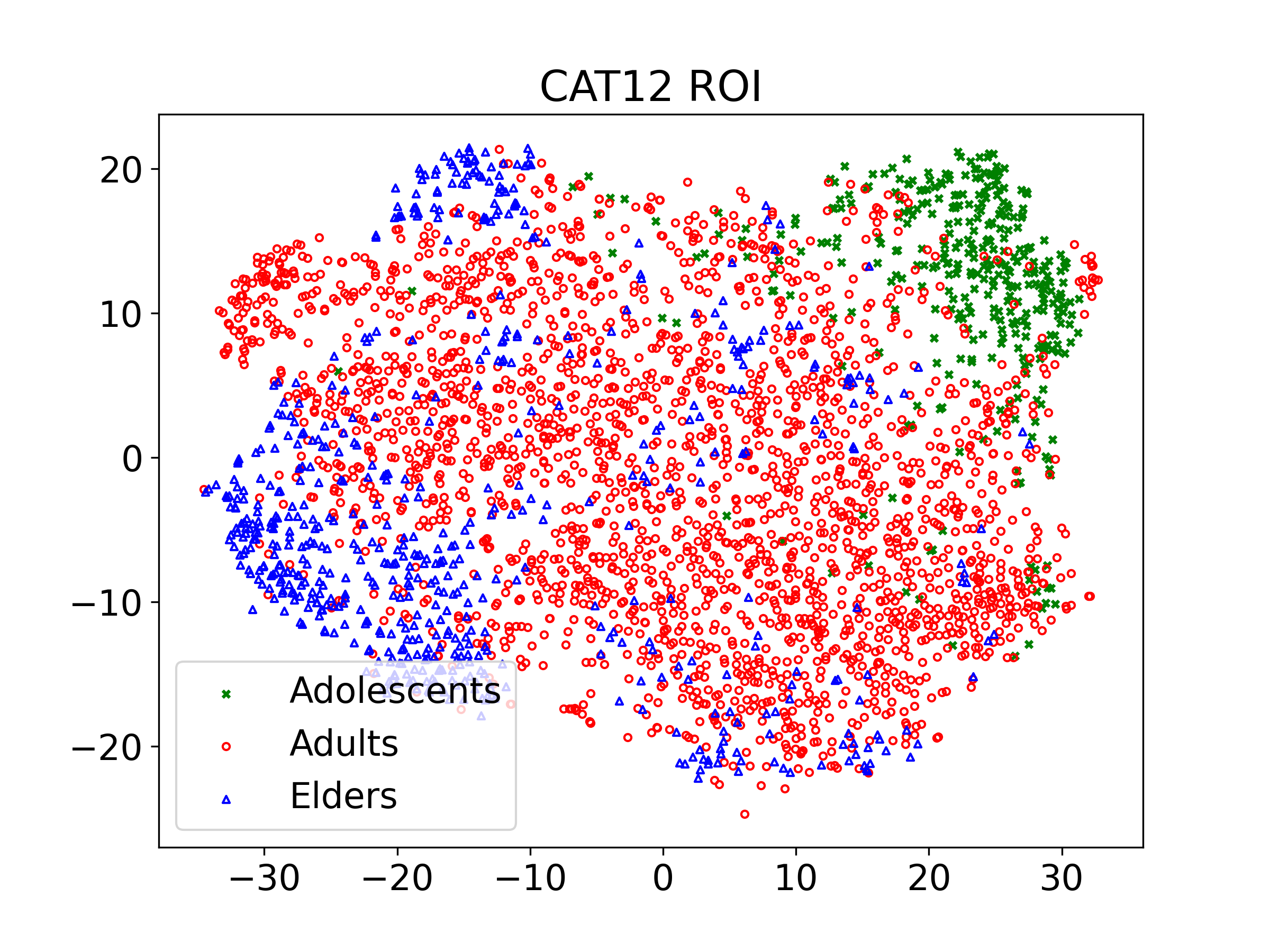}
        \label{fig:prob1_6_1}
    \end{minipage}%
    \begin{minipage}{0.48\textwidth}
        \centering
        \includegraphics[width=.88\linewidth]{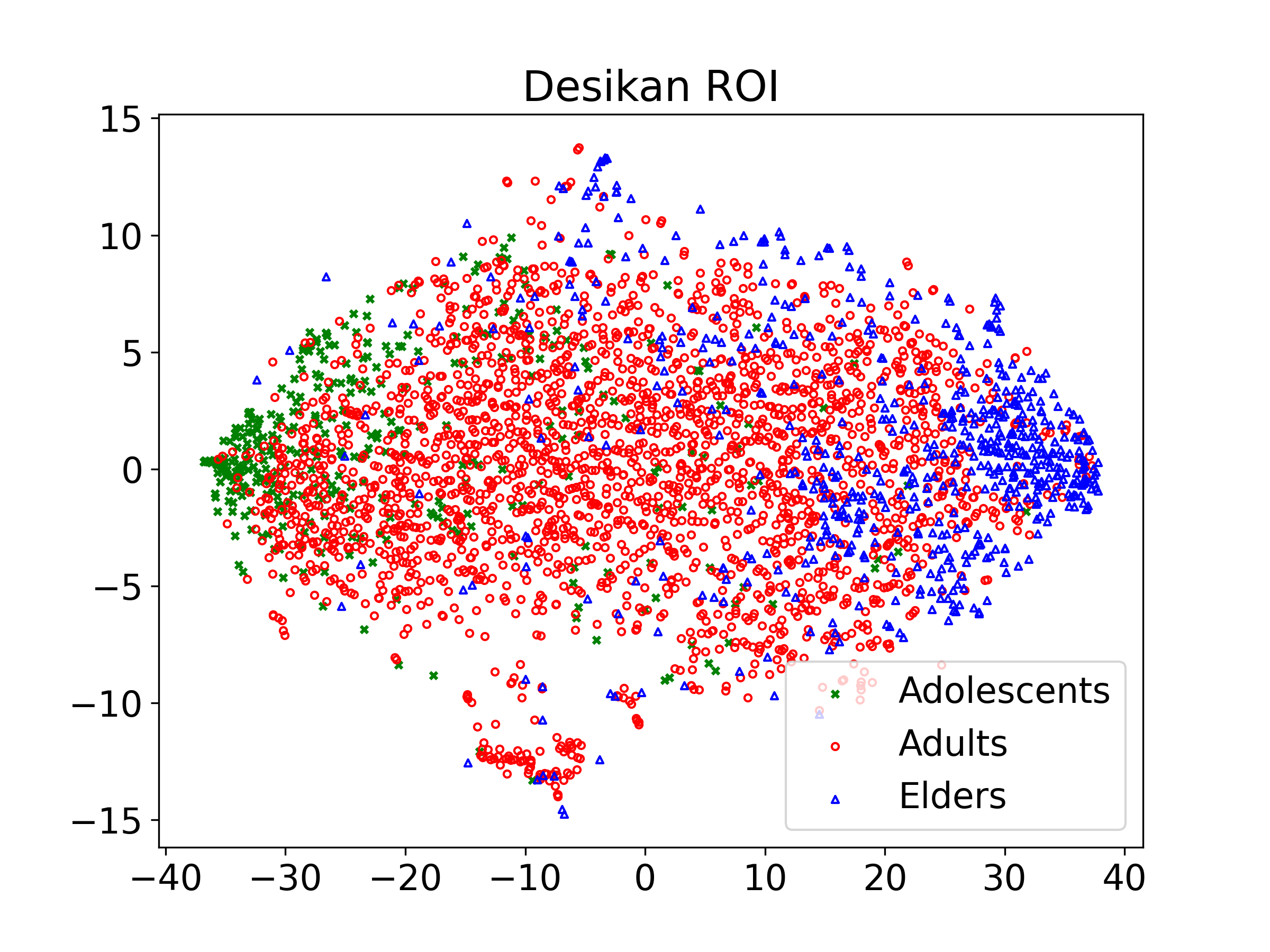}
        \label{fig:prob1_6_2}
    \end{minipage}\\%
    \par\vspace{-1.3\baselineskip}
    \begin{minipage}{0.48\textwidth}
        \centering
        \includegraphics[width=.88\linewidth]{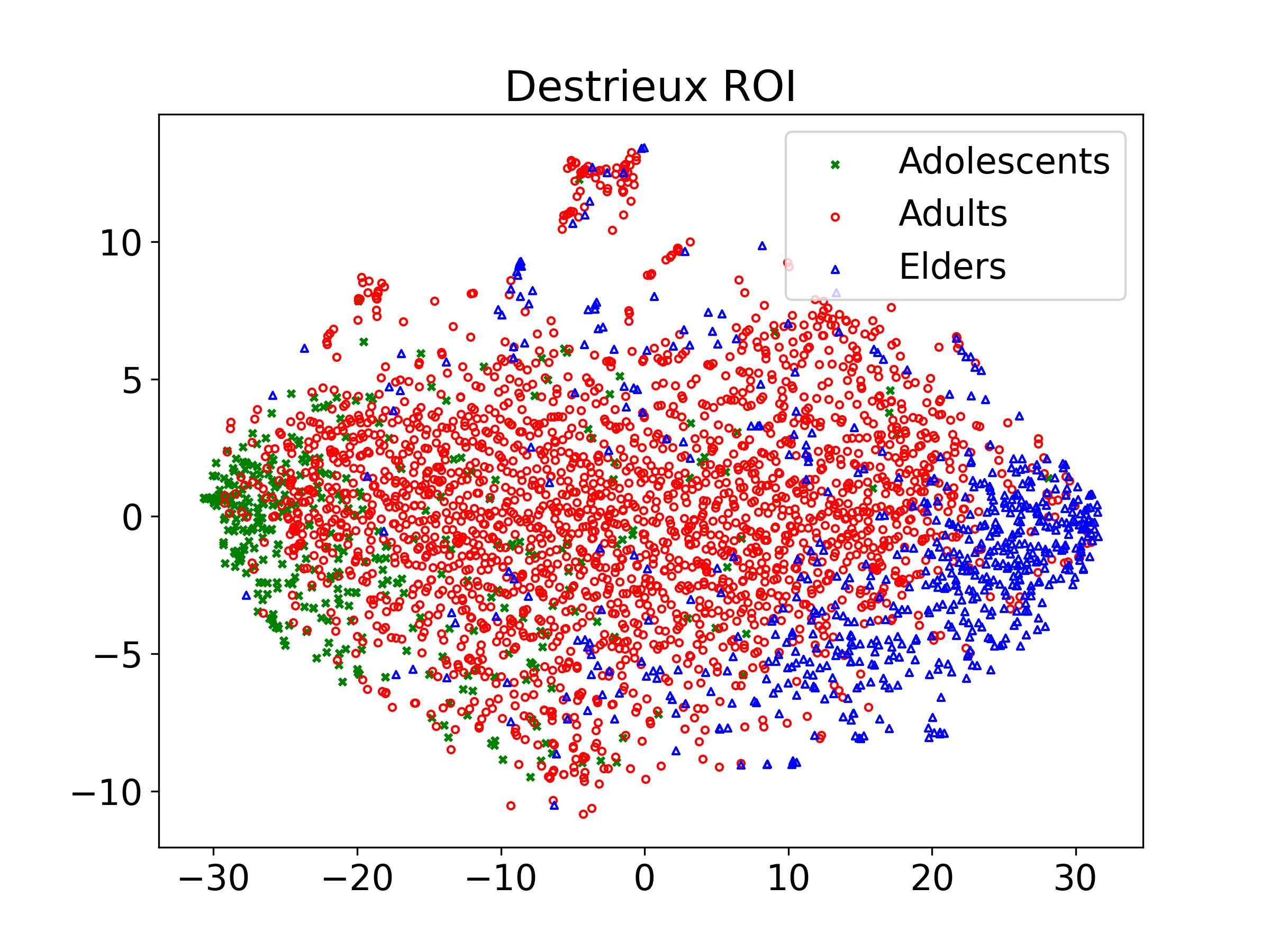}
        \label{fig:prob1_6_3}
    \end{minipage}%
     \begin{minipage}{.48\textwidth}
        \centering
        \includegraphics[width=.88\linewidth]{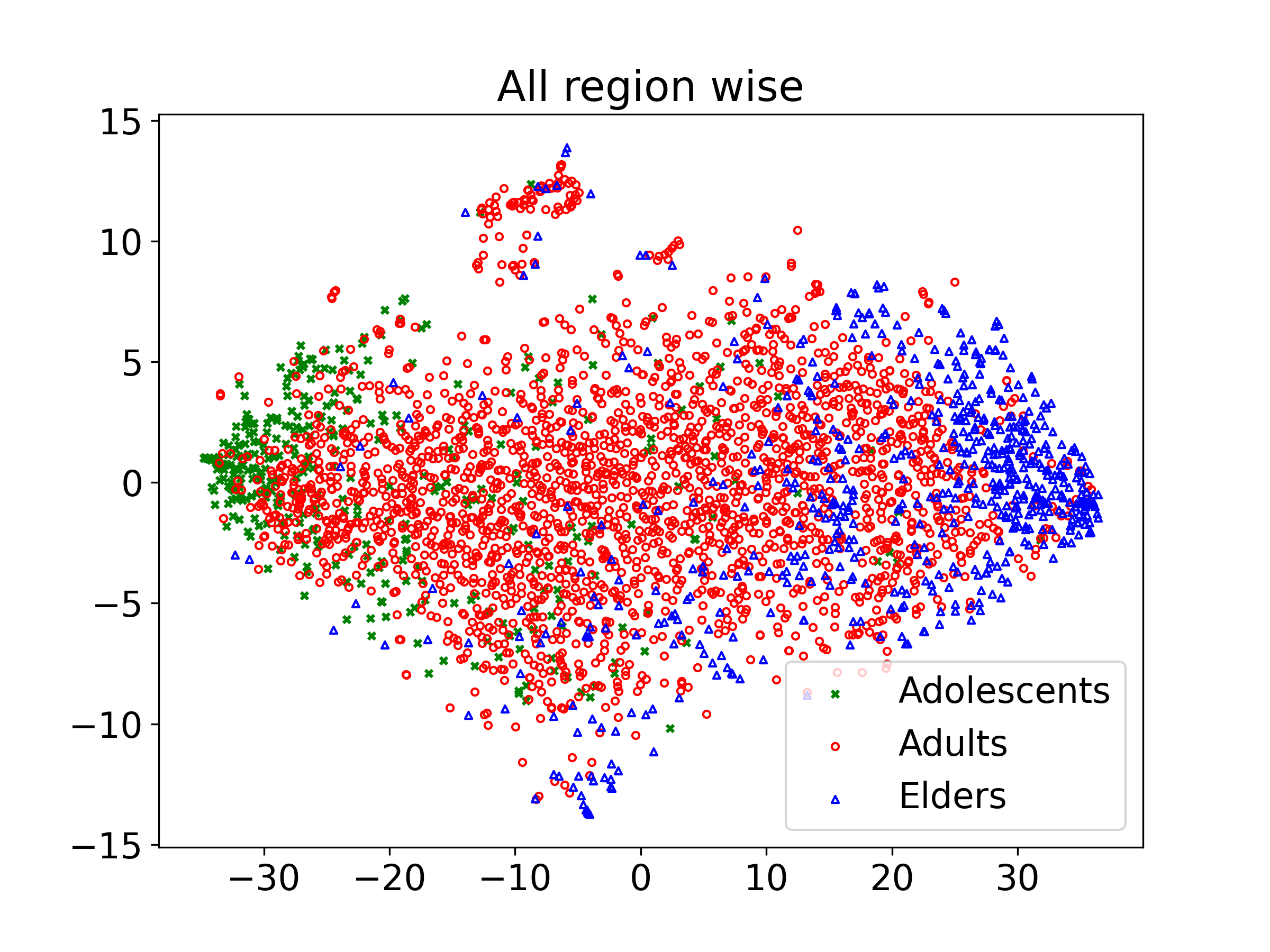}
        \label{fig:prob1_6_4}
    \end{minipage}
      \caption{$t$-SNE representation for different region-wise features of the three age groups.}
        \label{fig_tsne_features}  
\end{figure}

We represent the region-wise features of 3 age groups using $t$-SNE, as shown in Figure~\ref{fig_tsne_features}. We can observe that the ``Adolescents" class (in green color) shows a separate group and is far away from the ``Elders" class (in blue color). We also computed the correlation of features to test the statistical dependence with age (target variable). The features were min-max normalized to bring the values between 0 and 1 before training the regression model, $x_{scaled} = \nicefrac{x - x_{min}}{x_{max} - x_{min}}$.

We trained different regression models such as Support Vector Regression (SVR), Relevance Vector Regression (RVR), Linear Regression (LR), and Generalised Linear Model (GLM) to predict the brain age of the healthy test subjects. More formally, $Y^{pred}_m = \beta_1x_{m1} + \beta_2x_{m2} + . . . +  \beta_nx_{mn} + c$,	where $\beta_1$, $\beta_2$, and $\beta_n$ are the unknown parameters while $x_{m1}$, $x_{m1}$, and $x_{mn}$ are the selected brain MRI features.

\subsection{Working Algorithm}
Algorithm~\ref{brain_age_algo} describes our brain age estimation model. In line 1, we take data as input with their corresponding age vector. We train regression models on the input data of the $m$ training samples in lines 3 to 5 of the algorithm. Lastly, we validate the accuracy of our model using 10-fold cross-validation in line 6 and predict the brain ages of an unseen healthy test set in line 7 of the algorithm.

\begin{algorithm}[h!]
\caption{Brain age estimation model.}\label{brain_age_algo}
\small
\begin{algorithmic}[1]
\State\textbf{Data:} $X^{m \times n}$ $\gets$ Data matrix of $m$ subjects, $Y^{m \times 1}$ $\gets$ chronological ages of $m$ subjects.

\State\textbf{Result:} Let $Y^{pred}$ be the estimated ages of $m$ subjects.

\For{$m$ subjects}   \Comment{Model training}
    \State Train a regression model using the $X^{m \times n}$ with age labels $Y^{m \times 1}$
    \EndFor
\State Validate the prediction accuracy of the regression model
\Comment{Testing the regression model}
\State Estimate the brain age $Y^{pred}$ of HC test set

\end{algorithmic}
\end{algorithm}

\vspace{-1cm}
\section{Experimental Evaluation} \label{experiments}
In this section, we first present the existing baseline brain age estimation models. Then, we analyze the regression models used for developing BAE frameworks and choosing the best-performing algorithm using the evaluation metrics of MAE, RMSE, and R$^2$.

\subsection{Baseline Models}
We report comparison results of the proposed BAE model with the state-of-the-art BAE frameworks using T1-weighted MRI in Table~\ref{demographics}. The evaluation metrics for these brain age estimation models vary. For instance, some studies~\cite{BEHESHTI_BA_2022,LIU_2022_BA} have used MAE, RMSE, and R$^2$ to assess the performance of their models while others \cite{Cole2017_BA_CNN,Fujimoto_2017_bae_region,Jonsson_2019_BA} have used only MAE and R$^2$ to evaluate their BAE frameworks. The individual cohorts or subsets of healthy brain MRIs in the present study have previously been used in other BAE studies. For instance, \cite{BEHESHTI_BA_2022,Cole2017_BA_CNN,LIU_2022_BA} used the healthy brain MRI samples from IXI or ABIDE for brain age estimation that were also included in this study. Based on the age range, the number of training samples, and the evaluation metrics (MAE, RMSE, and R$^2$), our model outperforms the previously known brain age estimation models using T1-w MRIs as shown in Table~\ref{demographics}. 

\begin{table}[h!]
    \centering
    \resizebox{0.6\textwidth}{!}{
    \begin{tabular}{p{3cm}p{1cm}p{1.5cm}p{1cm}p{1cm}p{0.7cm}  }
     \toprule
     Study&\# HC &Age-range & MAE & RMSE &R$^2$ \\
    \midrule  \midrule    
     Beheshti et al.~\cite{BEHESHTI_BA_2022} & 788 & 18 - 94 &  4.7 & 6.13& 0.89\\
     Cole et al.~\cite{Cole2017_BA_CNN} &2001 & 18 - 90 & 4.16& - &0.95\\
     Aycheh et al.~\cite{Aycheh_2018_bae_region}& 2119 & 45 - 91 & 4.05 &  5.16 & - \\
     Jonsson et al.~\cite{Jonsson_2019_BA}&1264 & 20 - 86 & 3.85& - & 0.87  \\
     Liu et al.~\cite{LIU_2022_BA} & 2501 & 20 - 94 &  3.73&4.81 & 0.95\\
     Baecker et al.~\cite{Baecker_2021_bae_reg} & 10824 & 47 - 73 & 3.7 & 4.65 & - \\
     Fujimoto et al.~\cite{Fujimoto_2017_bae_region}& 1099 & 20 - 75 &3.59 & - &0.95 \\
     \textbf{Proposed model} & \textbf{3965} & \textbf{ 6 - 86} & \textbf{3.25} & \textbf{4.72} & \textbf{0.90} \\
     \bottomrule
    \end{tabular} 
    }\medskip
    \caption{A comparison of the studies conducted for brain age estimation using T1-weighted MRI and our proposed model. }
    \label{demographics}
 \end{table}

\vspace{-1cm}
\subsection{Regression Algorithms}
We compare the performance of each regression model (LR, SVR, RVR, and GLM) using different region-wise features in Table~\ref{tbl_regression}. Our analysis showed that GLM outperformed the other models, achieving lower mean absolute error (MAE) with the FreeSurfer Desikan and Destrieux ROI features. In contrast, RVR achieved better performance with the CAT12 ROI features. However, GLM produced the lowest MAE for all features combined, i.e., all region-wise ({\small MAE} = 3.25 years).

\section{Results and Discussion} \label{results}
In this section, we present the results of our proposed model and compare them to existing baselines. We begin by showing the regression results using different evaluation metrics and methods in Table~\ref{tbl_regression}. Specifically, we trained the regression algorithms (LR, SVR, RVR, and GLM) on each feature set and computed the performance evaluation metrics after tuning the hyperparameters. We then selected the best-performing model based on lower mean absolute error (MAE), root mean squared error (RMSE), and higher R$^2$ score.
\begin{table}[h!]
    \centering
    \footnotesize
    \resizebox{0.45\textwidth}{!}{
    \begin{tabular}{p{2.5cm}lp{1.1cm}p{1.1cm}p{1.1cm}}
    \toprule
        \multirow{1}{*}{Features} & \multirow{1}{1.5cm}{Algo.} & \multirow{1}{1.3cm}{MAE $\downarrow$} & \multirow{1}{1.3cm}{RMSE $\downarrow$} & \multirow{1}{1.6cm}{$R^2$ $\uparrow$} \\
        \midrule \midrule
        \multirow{3}{3cm}{CAT12 ROI}
        & LR & 4.24 & 5.73 & 0.85 	\\
        & SVR & 4.26 & 5.85 & 0.85     \\
        & \textbf{RVR} & \textbf{3.94} & \textbf{5.32} & \textbf{0.86}  \\
        & GLM & 4.02 &  5.88 & 0.84  \\
        \cmidrule{2-5}
        \multirow{3}{3cm}{Desikan ROI}
        & LR & 5.37 & 7.11 & 0.78	\\
        & SVR & 5.28 & 7.14 & 0.77     \\
        & RVR & 5.87 & 10.76 & 0.46  \\
        & \textbf{GLM} & \textbf{4.23} &  \textbf{6.4} & \textbf{0.81}  \\
        \cmidrule{2-5}
        \multirow{3}{3cm}{Destrieux ROI}
        & LR & 5.31 & 7.03 & 0.79 	\\
        & SVR & 5.12 & 6.99 & 0.80     \\
        & RVR & 5.50 &  11.31 & 0.41 \\
        & \textbf{GLM} & \textbf{3.9} &  \textbf{5.81} & \textbf{0.84}  \\
        \cmidrule{2-5}
        \multirow{3}{2.5cm}{CAT12 ROI + Desikan ROI}
        & LR & 4.21 & 5.54 & 0.86 	\\
        & SVR & 4.12 & 5.51 & 0.86     \\
        & RVR & 3.56 &  4.92 & 0.89  \\
        & \textbf{GLM} & \textbf{3.4} &  \textbf{5.02} & \textbf{0.88}  \\
        \cmidrule{2-5}
        \multirow{3}{3cm}{CAT12 ROI + Destrieux ROI}
        & LR & 4.61 & 5.97 & 0.85 	\\
        & SVR & 4.42 & 5.65 & 0.87     \\
        & RVR & 3.74 &  5.00 & 0.88  \\
        & \textbf{GLM} & \textbf{3.33} &  \textbf{4.87} & \textbf{0.89}  \\
        \cmidrule{2-5}
        \multirow{3}{3cm}{Desikan ROI + Destrieux ROI}
        & LR & 5.74 & 7.58 & 0.74 	\\
        & SVR & 5.17 & 6.82 & 0.80     \\
        & RVR & 5.54 &  11.08 & 0.43  \\
        & \textbf{GLM} & \textbf{3.77} &  \textbf{5.58} & \textbf{0.86}  \\
        \cmidrule{2-5}
        \multirow{3}{3cm}{All region wise}
        & LR & 5.31 & 6.62 & 0.79 	\\
        & SVR & 4.49 & 5.73 & 0.85     \\
        & RVR & 3.37 &  4.91 & 0.89  \\
        & \textbf{GLM} & \textbf{3.25} &  \textbf{4.73} & \textbf{0.90}  \\
    \bottomrule
         \end{tabular}
    }\medskip
    \caption{Regression results for different MRI-derived region-wise features (LR: Linear Regression, SVR: Support Vector Regression, RVR: Relevance Vector Regression, GLM: Generalized Linear Model).}
    \label{tbl_regression}
\end{table}

\begin{figure}[h!]
     \centering    
     \begin{minipage}{0.25\textwidth}
        \centering
        \includegraphics[width=0.98\linewidth]{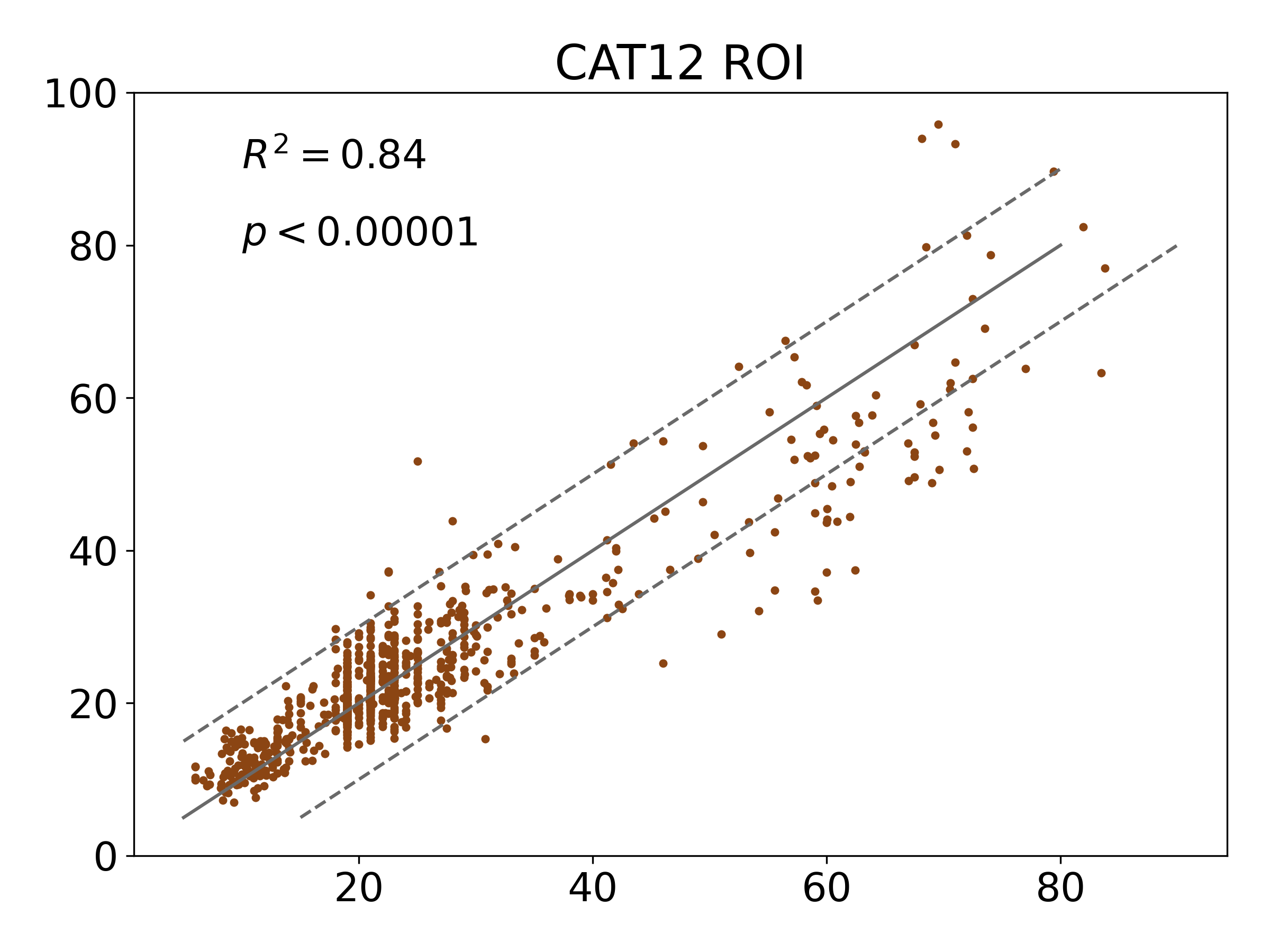}
        \label{fig:prob1_4_1}
    \end{minipage}%
    \begin{minipage}{0.25\textwidth}
        \centering
        \includegraphics[width=0.98\linewidth]{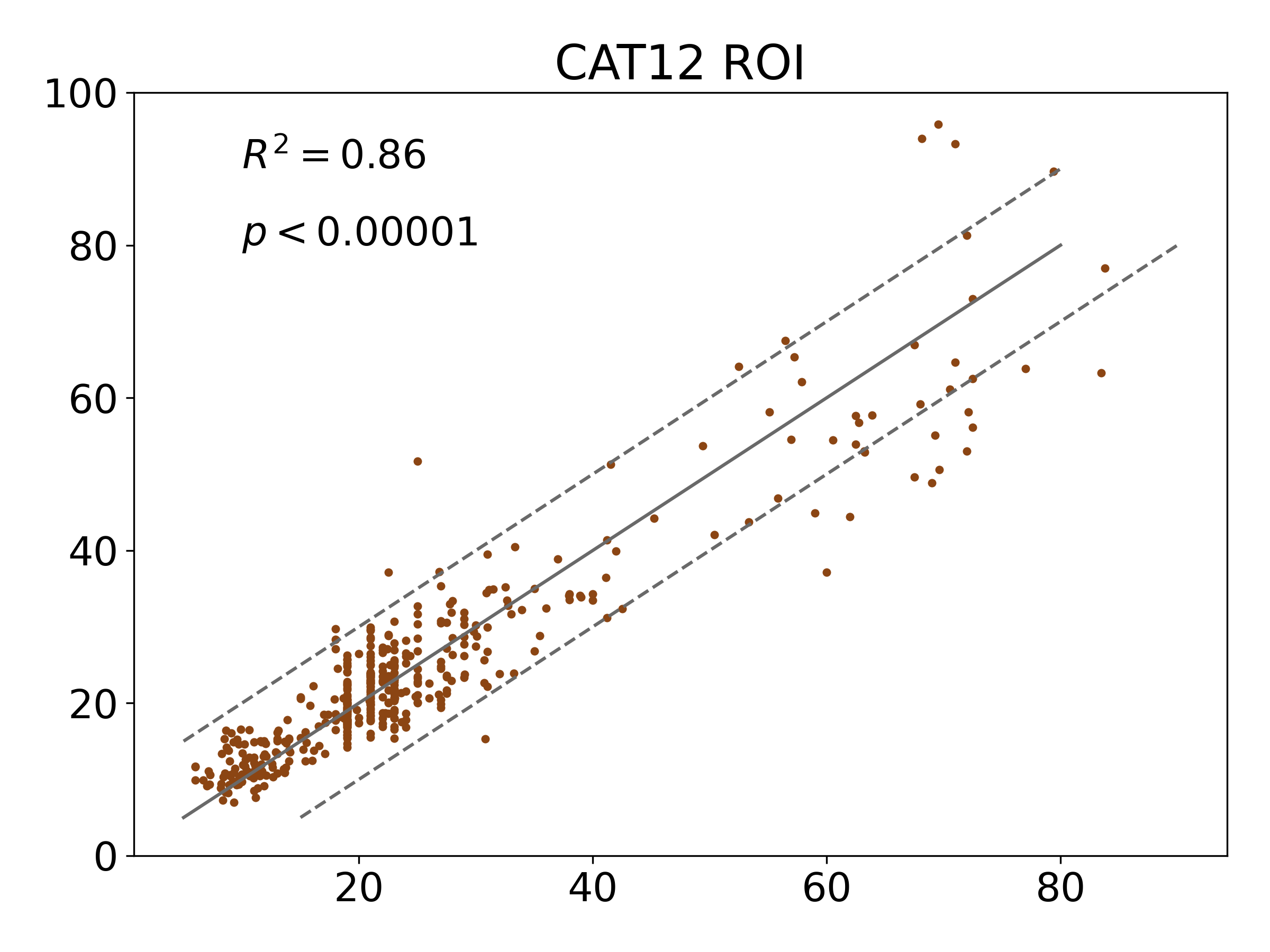}
        \label{fig:prob1_4_2}
    \end{minipage}%
    \begin{minipage}{0.25\textwidth}
        \centering
        \includegraphics[width=0.98\linewidth]{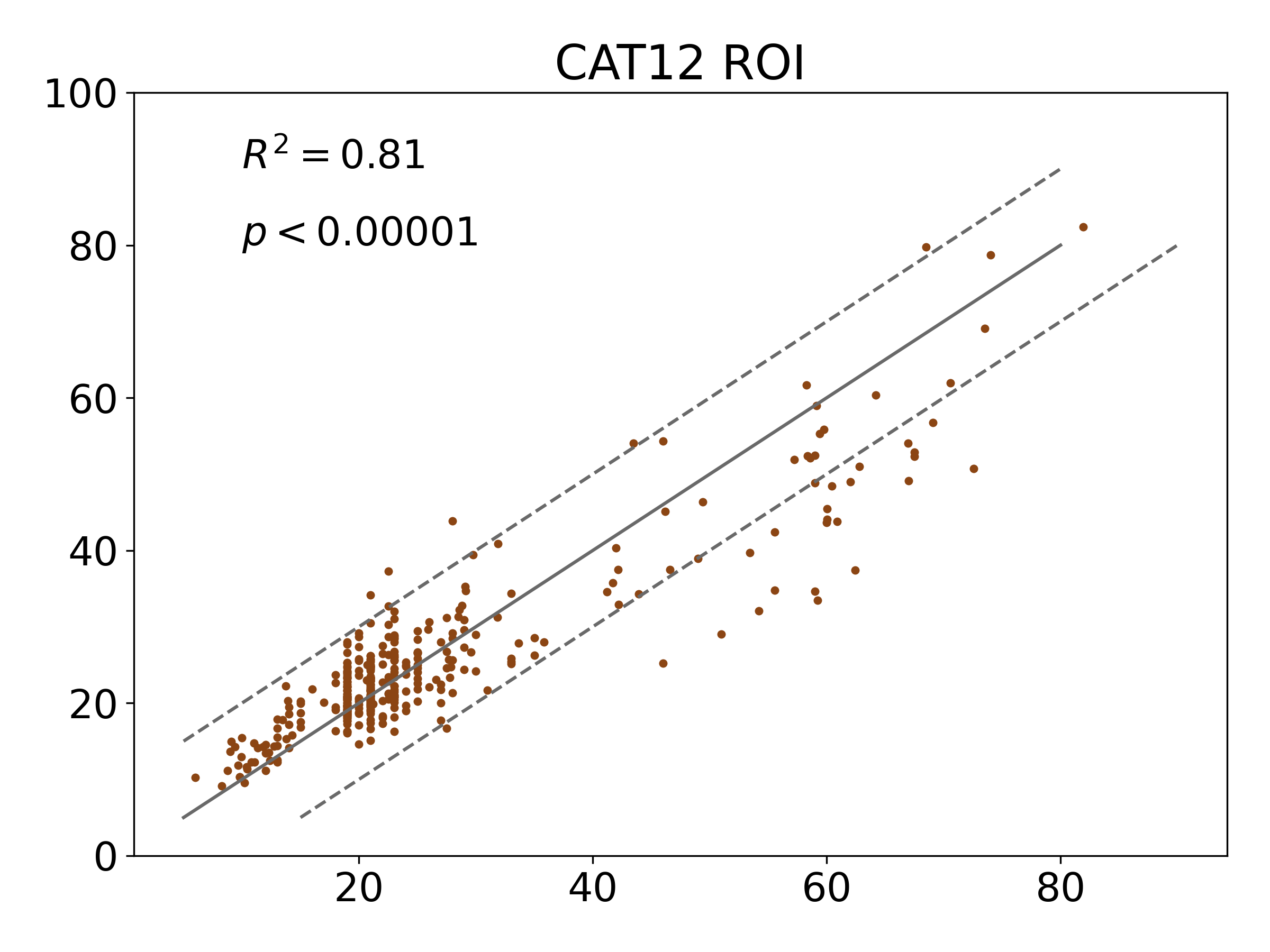}
        \label{fig:prob1_4_3}
    \end{minipage}
    \par\vspace{-1.3\baselineskip}
     \begin{minipage}{0.25\textwidth}
        \centering
        \includegraphics[width=0.98\linewidth]{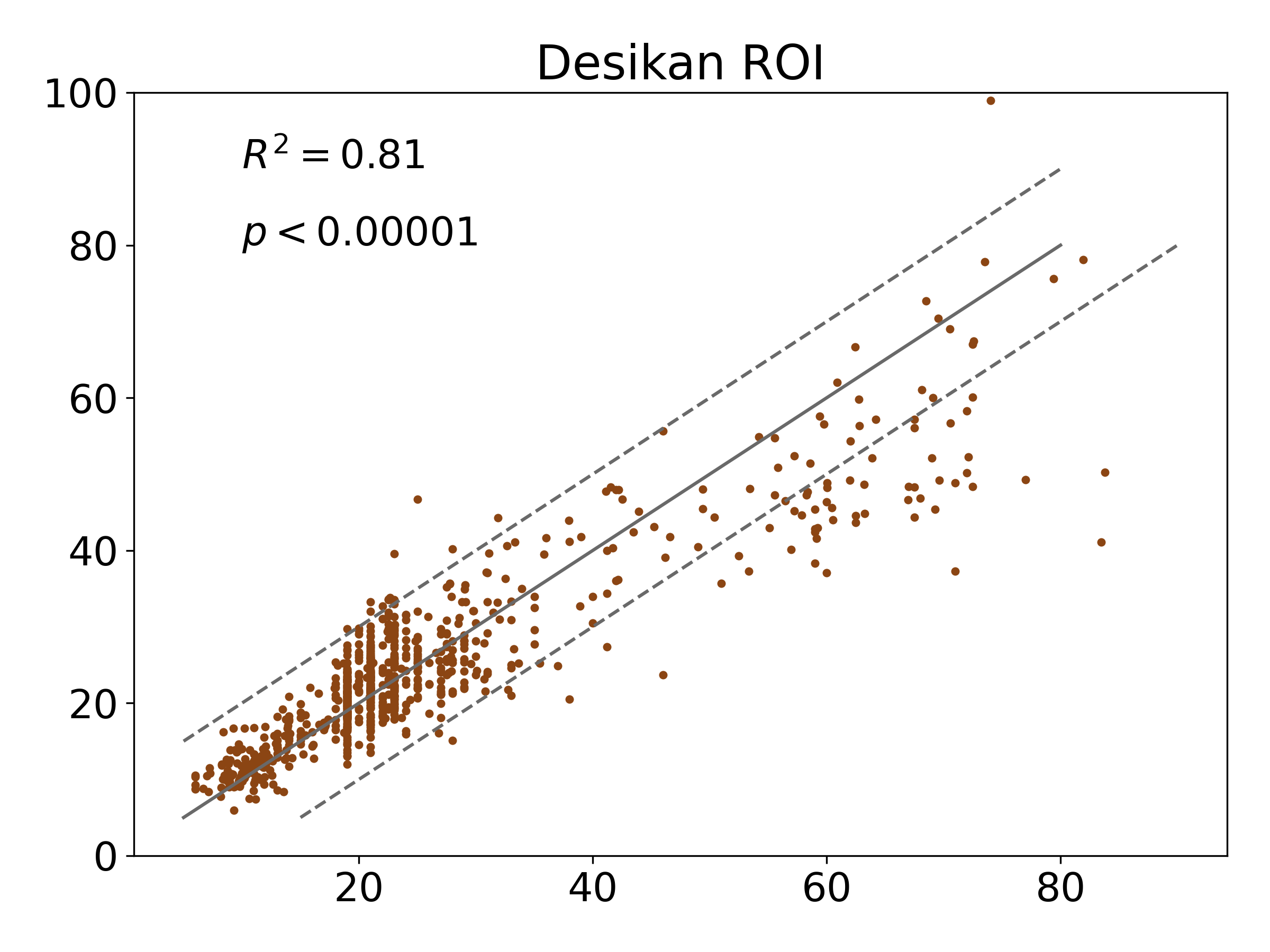}
        \label{fig:prob1_4_4}
    \end{minipage}%
    \begin{minipage}{0.25\textwidth}
        \centering
        \includegraphics[width=0.98\linewidth]{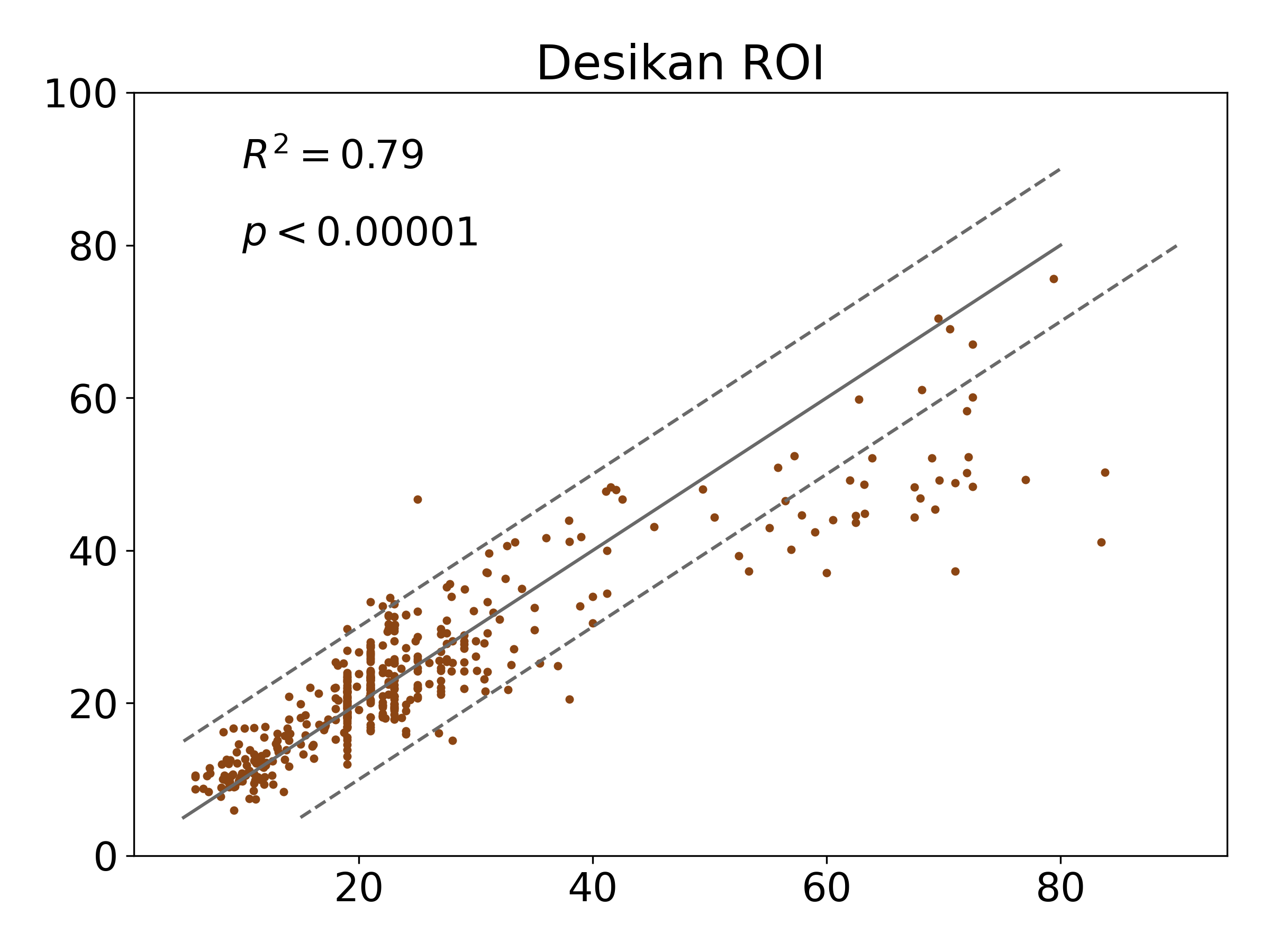}
        \label{fig:prob1_4_5}
    \end{minipage}%
    \begin{minipage}{0.25\textwidth}
        \centering
        \includegraphics[width=0.98\linewidth]{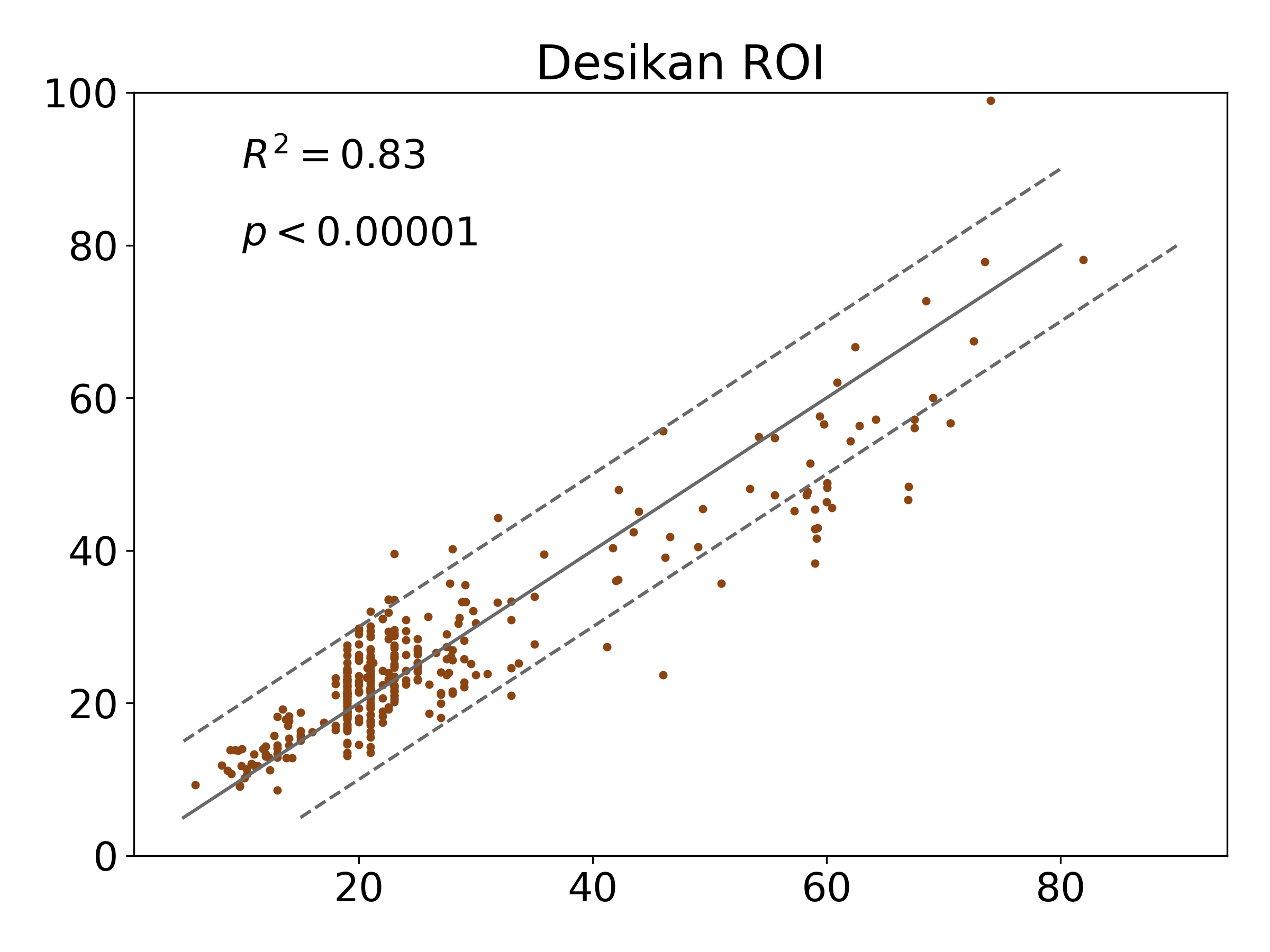}
        \label{fig:prob1_4_6}
    \end{minipage}
    \par\vspace{-1.3\baselineskip}
    \begin{minipage}{0.25\textwidth}
        \centering
        \includegraphics[width=0.98\linewidth]{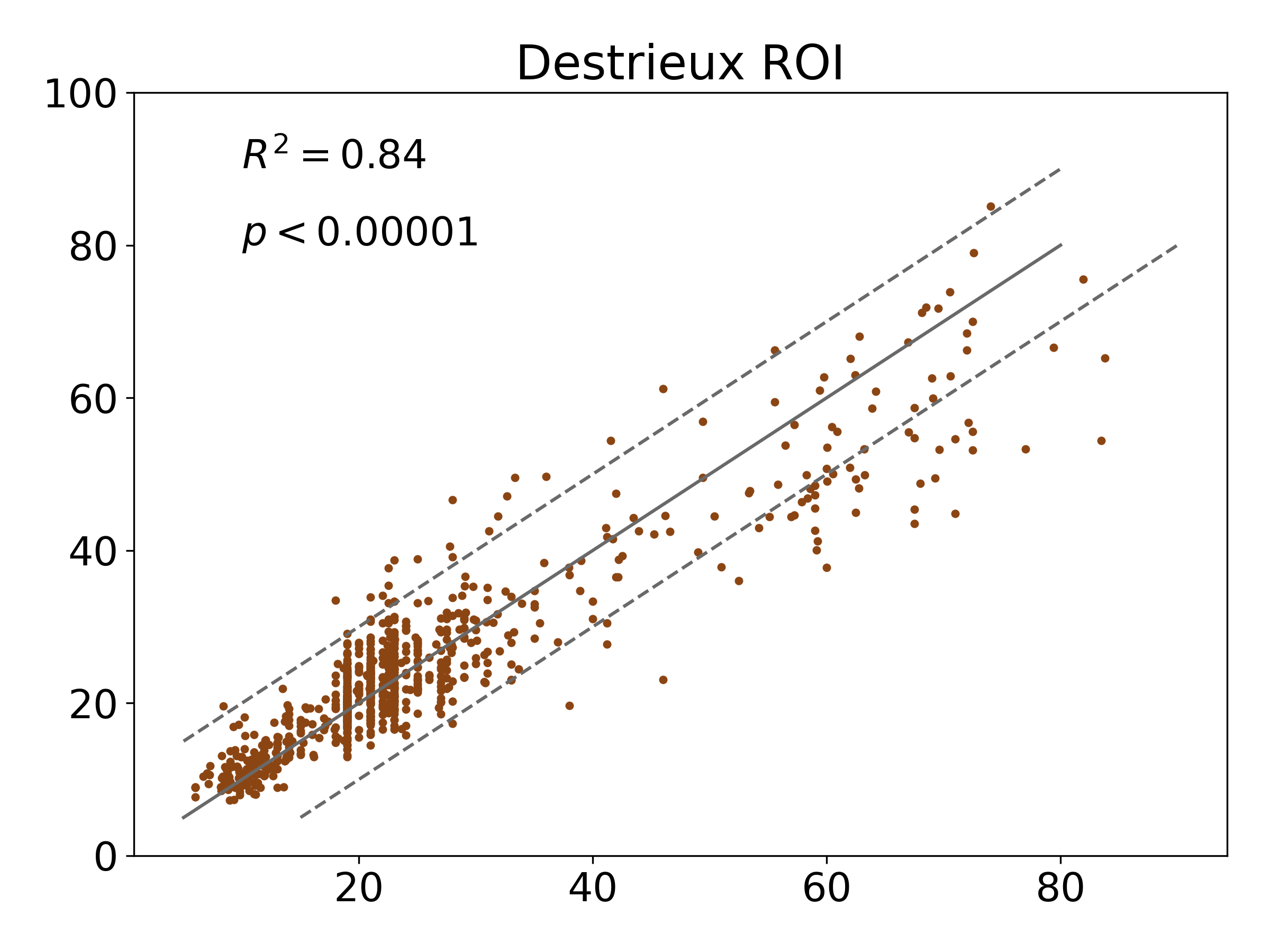}
        \label{fig:prob1_4_7}
    \end{minipage}
     \begin{minipage}{0.25\textwidth}
        \centering
        \includegraphics[width=0.98\linewidth]{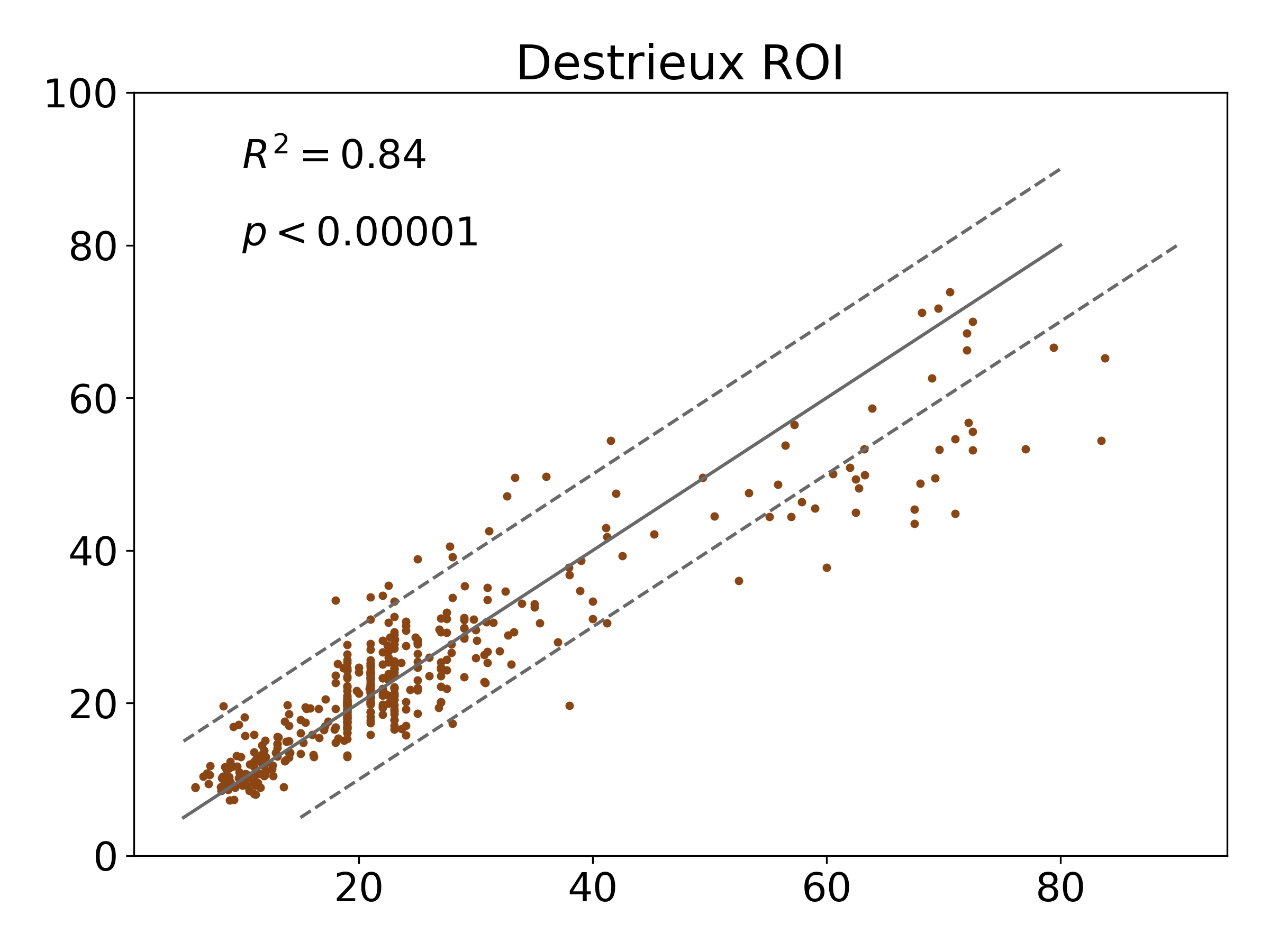}
        \label{fig:prob1_4_8}
    \end{minipage}%
    \begin{minipage}{0.25\textwidth}
        \centering
        \includegraphics[width=0.98\linewidth]{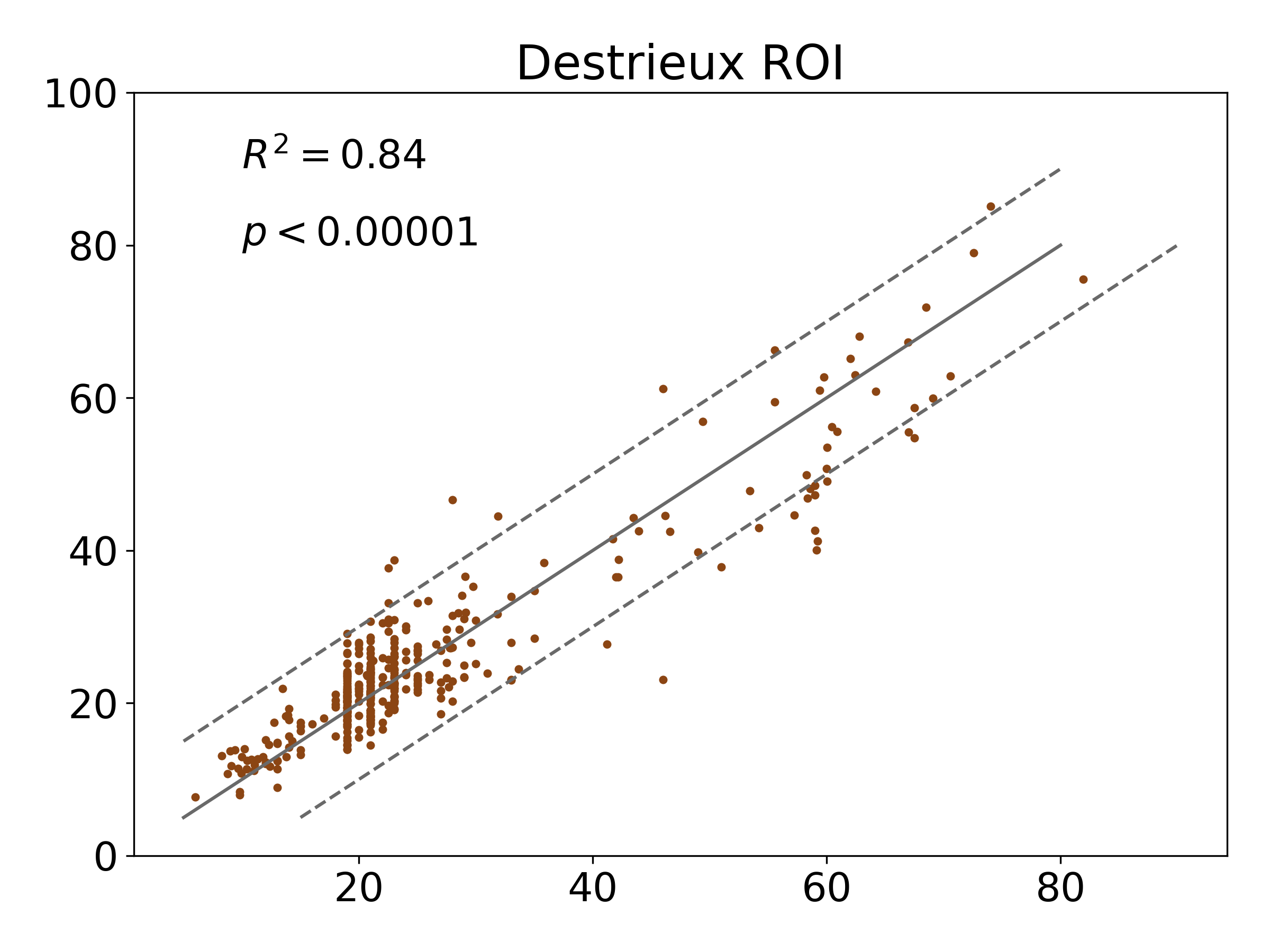}
        \label{fig:prob1_4_9}
    \end{minipage}
    \par\vspace{-1.3\baselineskip}
    \begin{minipage}{0.25\textwidth}
        \centering
        \includegraphics[width=0.98\linewidth]{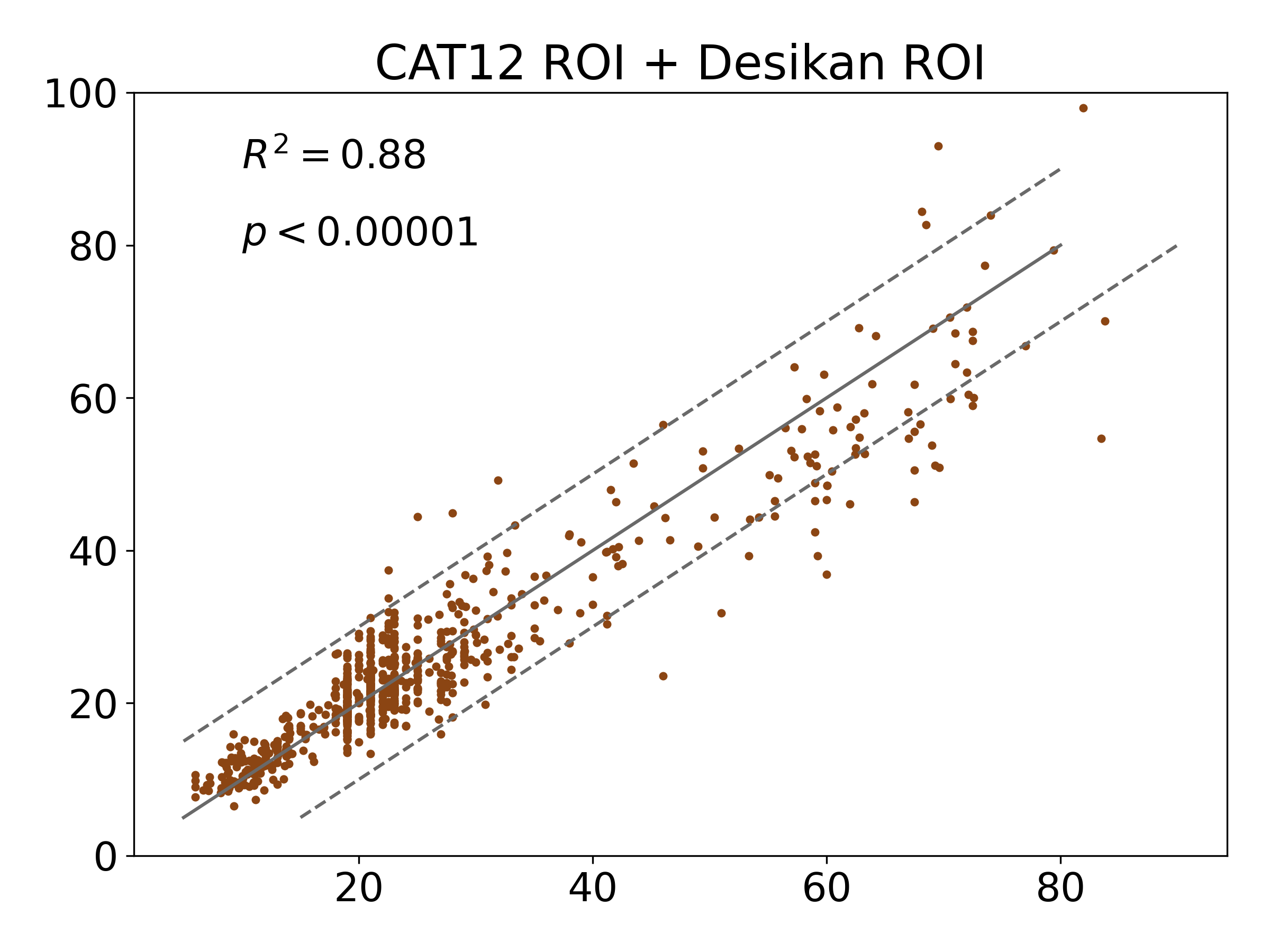}
        \label{fig:prob1_4_10}
    \end{minipage}%
     \begin{minipage}{0.25\textwidth}
        \centering
        \includegraphics[width=0.98\linewidth]{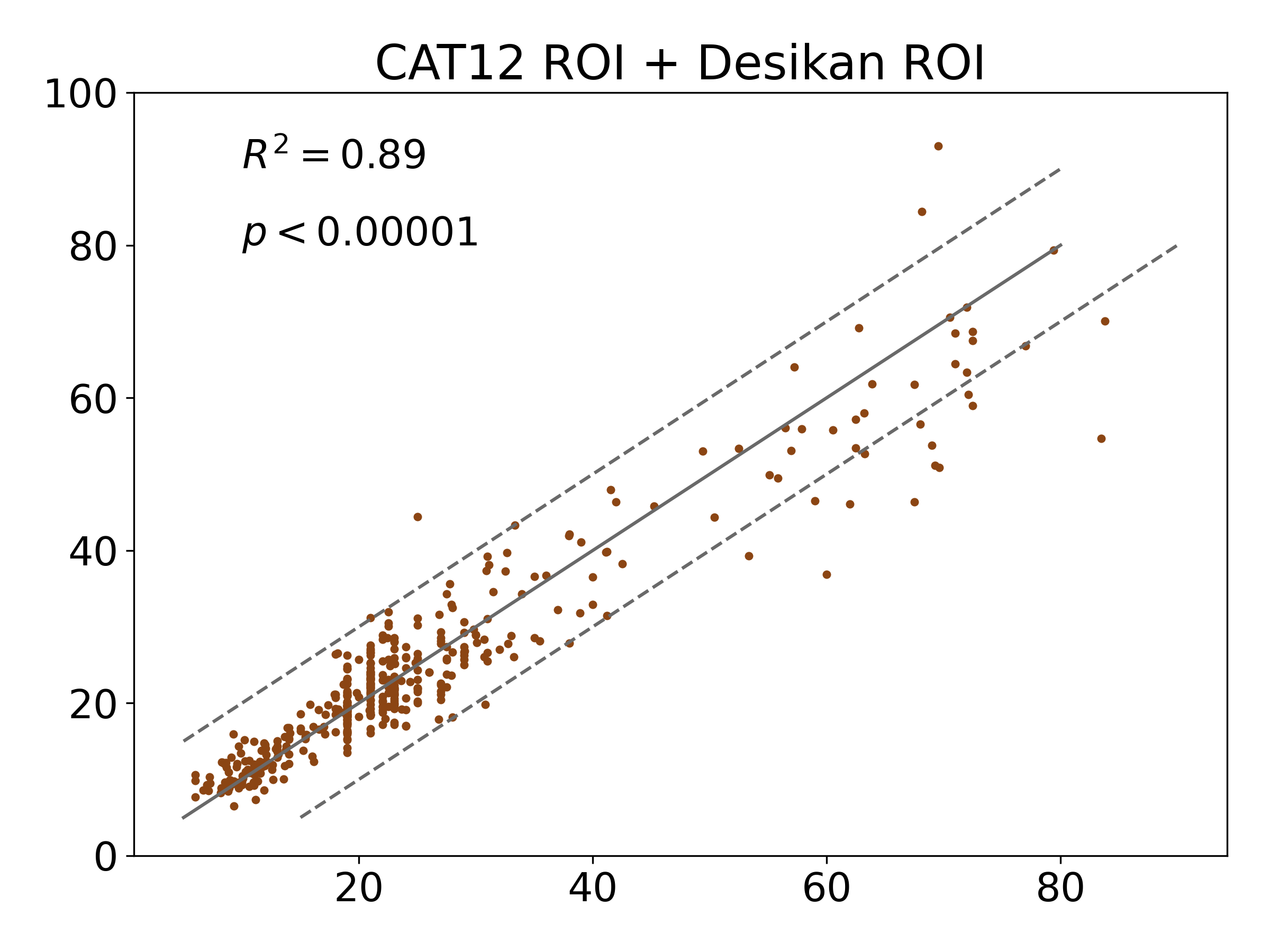}
        \label{fig:prob1_4_11}
    \end{minipage}%
    \begin{minipage}{0.25\textwidth}
        \centering
        \includegraphics[width=0.98\linewidth]{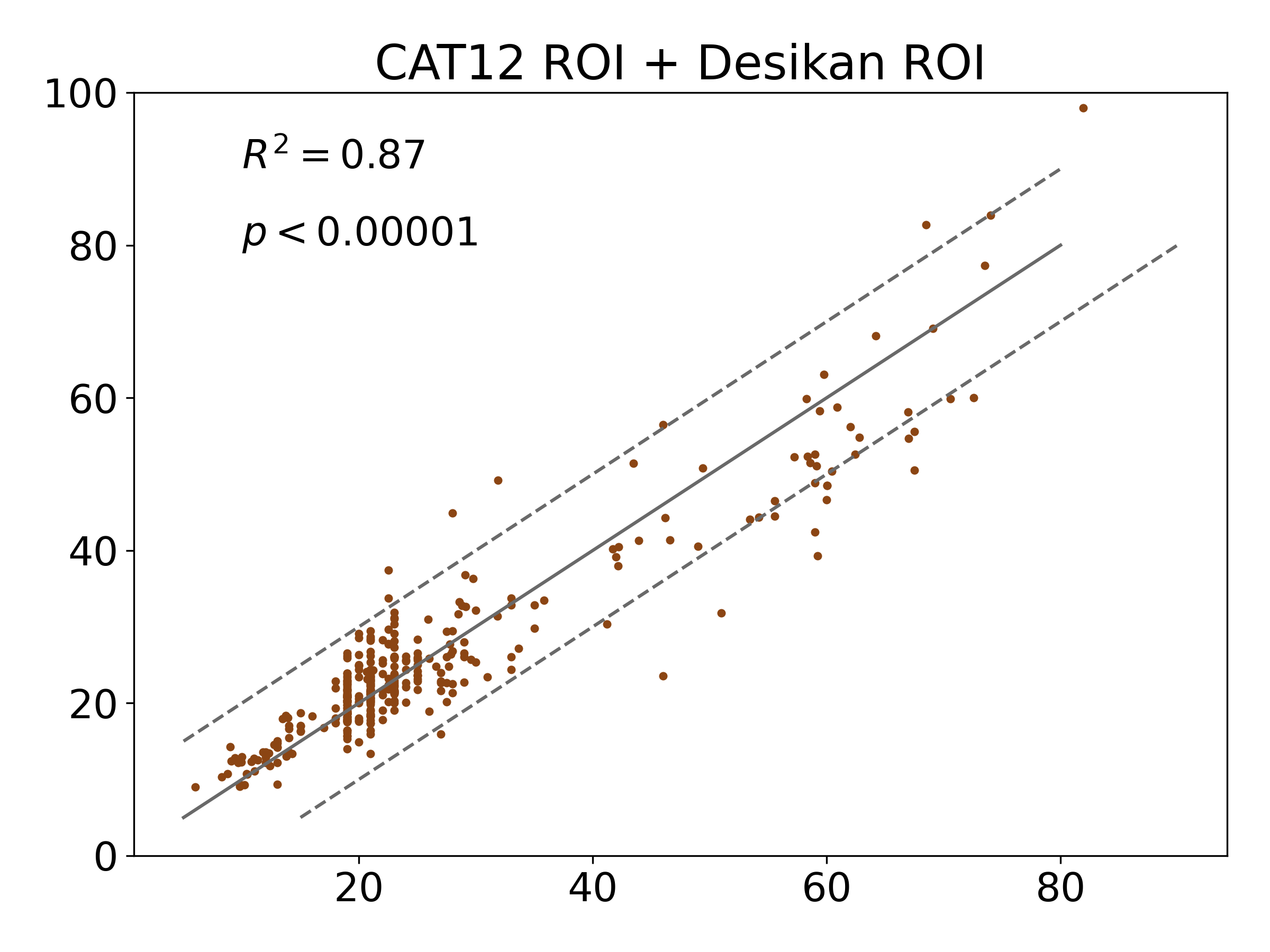}
        \label{fig:prob1_4_12}
    \end{minipage}
    \par\vspace{-1.3\baselineskip}
    \begin{minipage}{0.25\textwidth}
        \centering
        \includegraphics[width=0.98\linewidth]{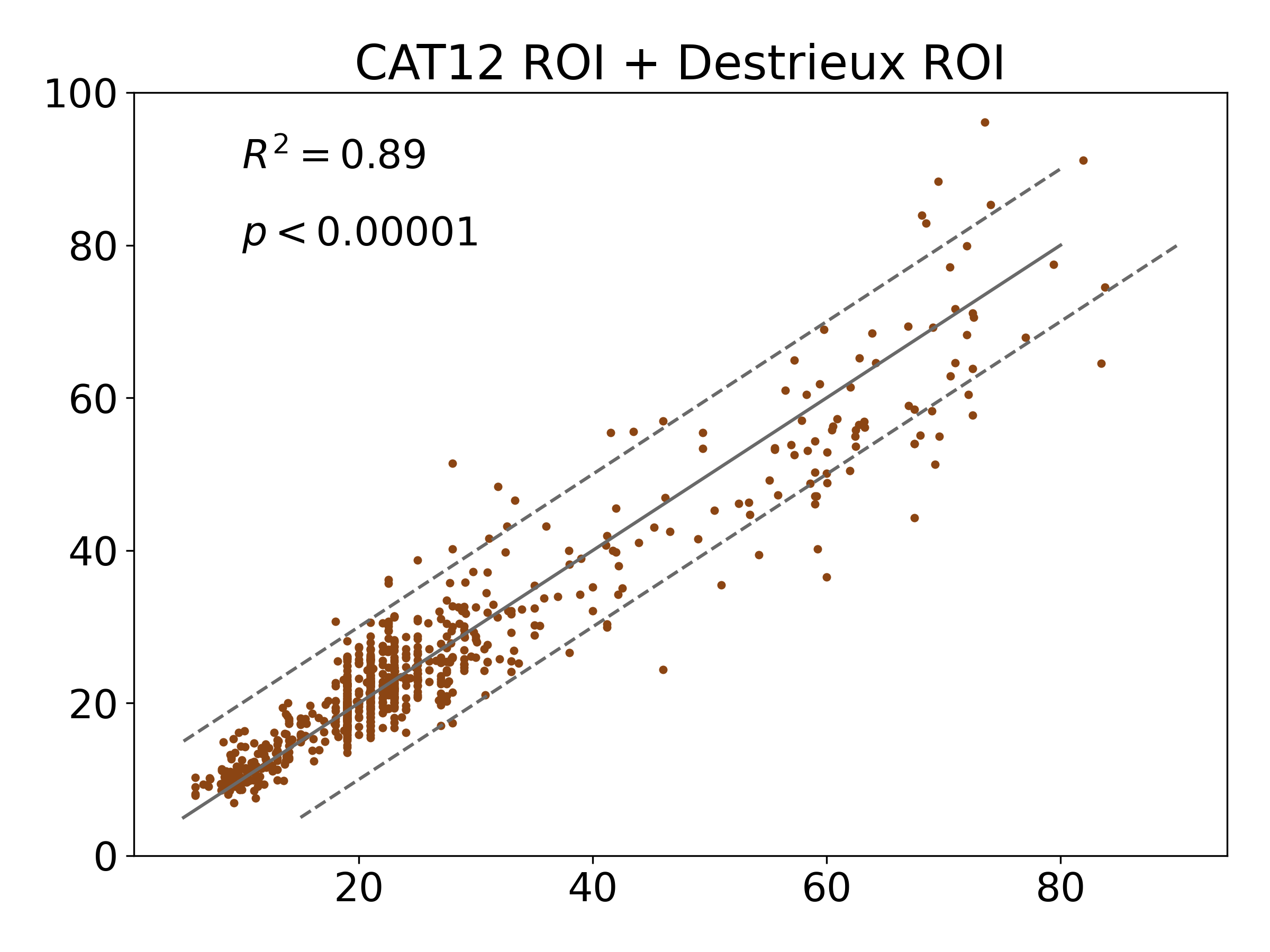}
        \label{fig:prob1_4_13}
    \end{minipage}%
    \begin{minipage}{0.25\textwidth}
        \centering
        \includegraphics[width=0.98\linewidth]{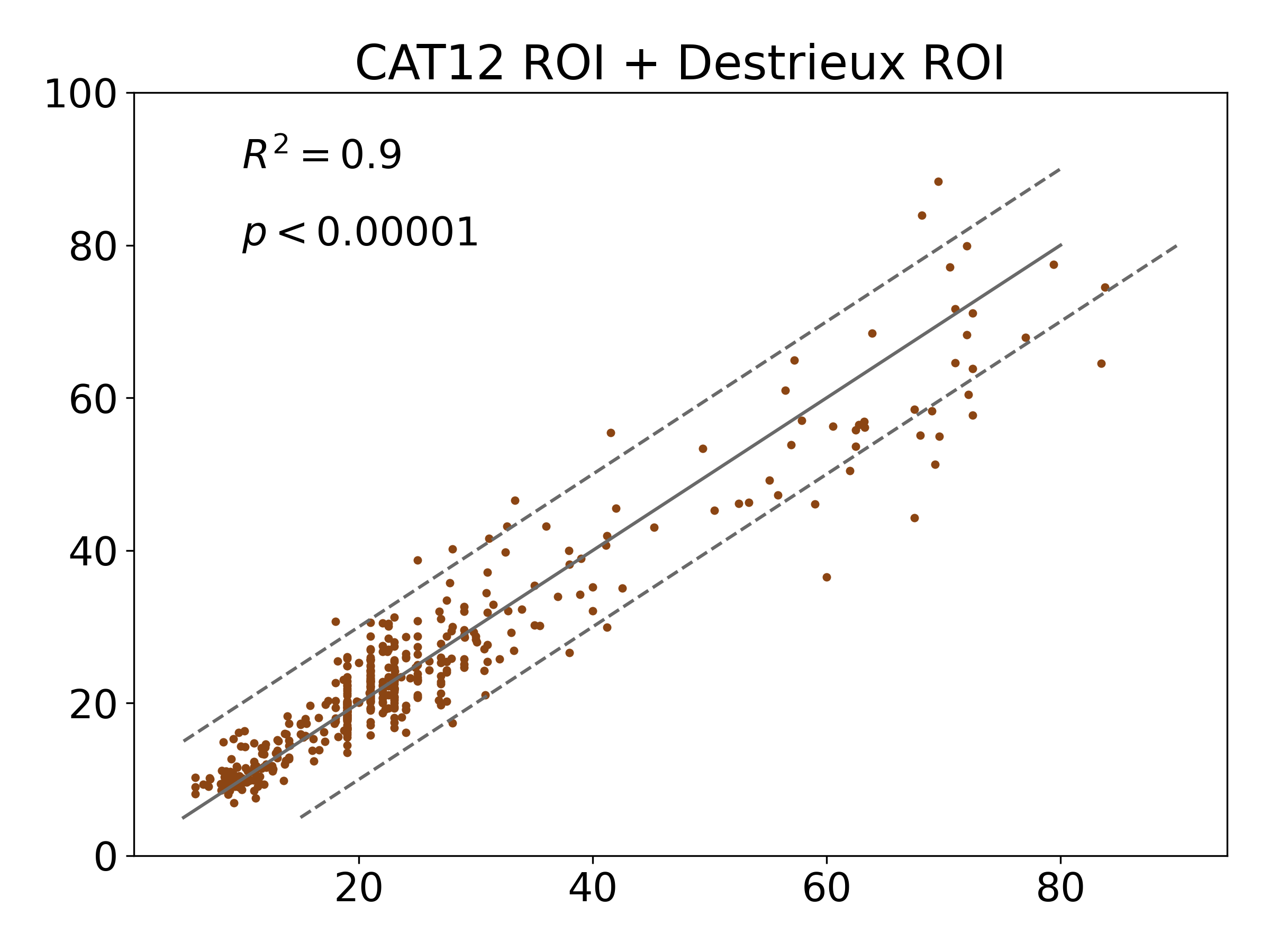}
        \label{fig:prob1_4_14}
    \end{minipage} %
     \begin{minipage}{0.25\textwidth}
        \centering
        \includegraphics[width=0.98\linewidth]{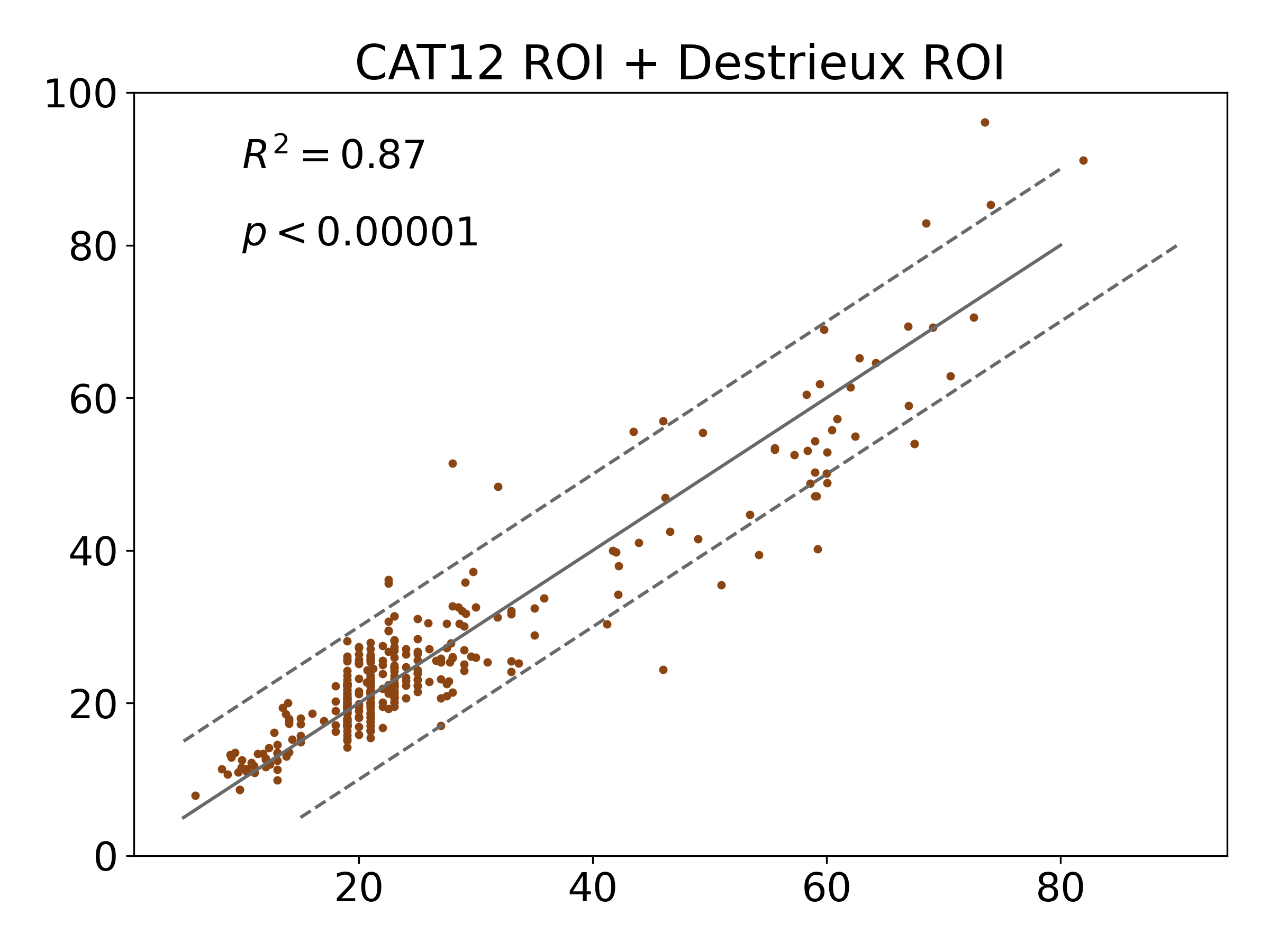}
        \label{fig:prob1_4_15}
    \end{minipage}
    \par\vspace{-1.3\baselineskip}
    \begin{minipage}{0.25\textwidth}
        \centering
        \includegraphics[width=0.98\linewidth]{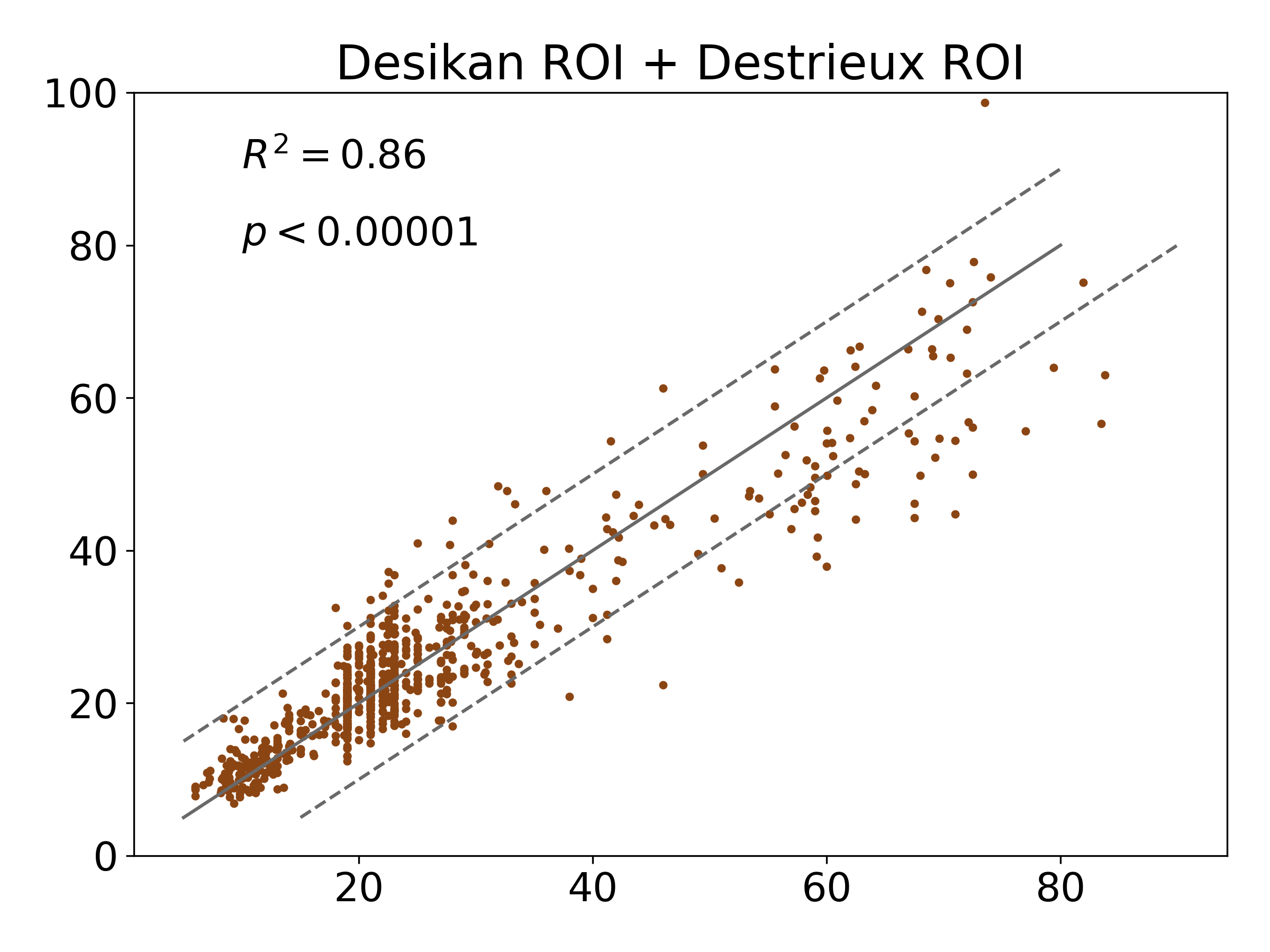}
        \label{fig:prob1_4_16}
    \end{minipage}%
    \begin{minipage}{0.25\textwidth}
        \centering
        \includegraphics[width=0.98\linewidth]{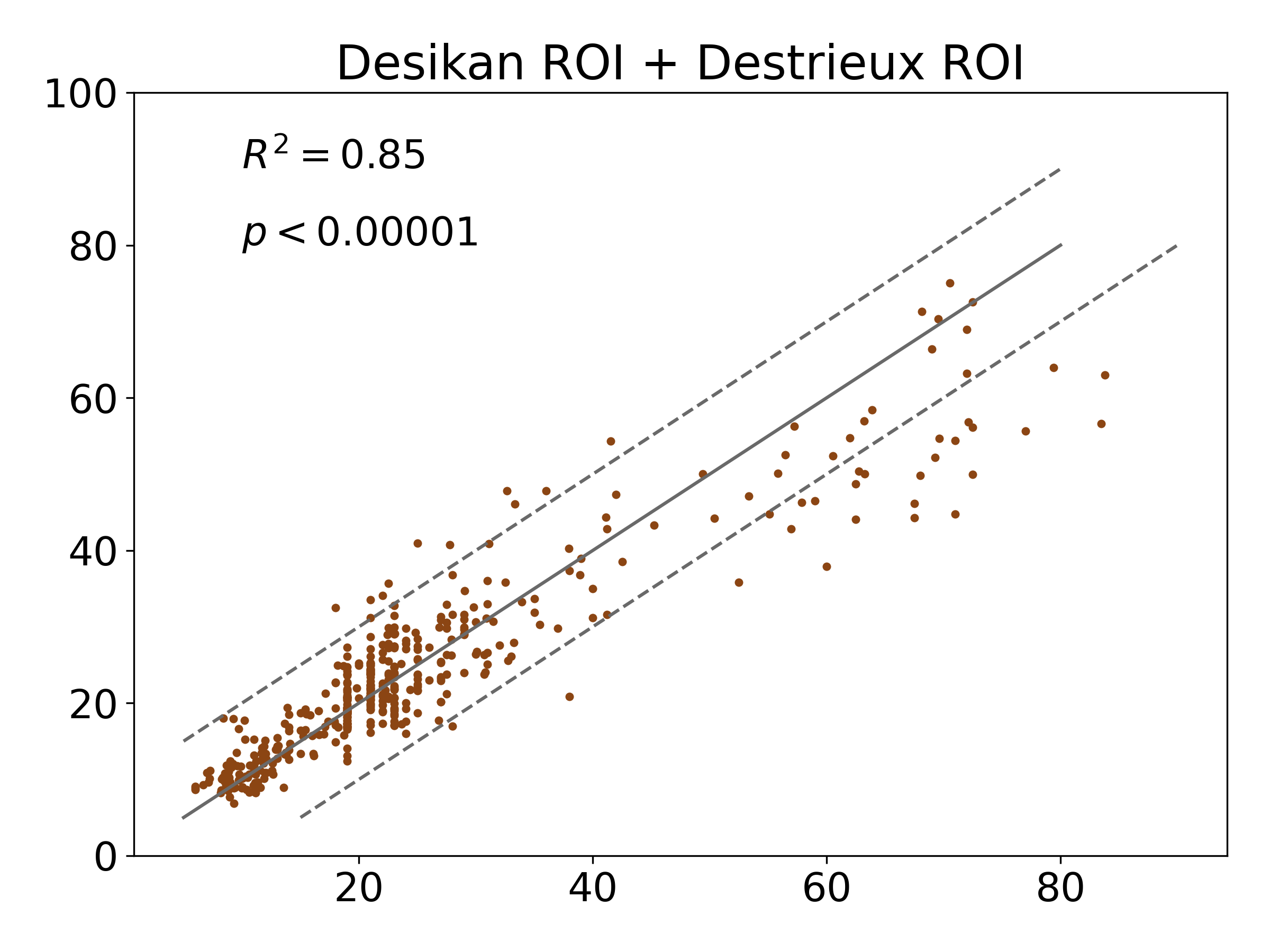}
        \label{fig:prob1_4_17}
    \end{minipage}%
     \begin{minipage}{0.25\textwidth}
        \centering
        \includegraphics[width=0.98\linewidth]{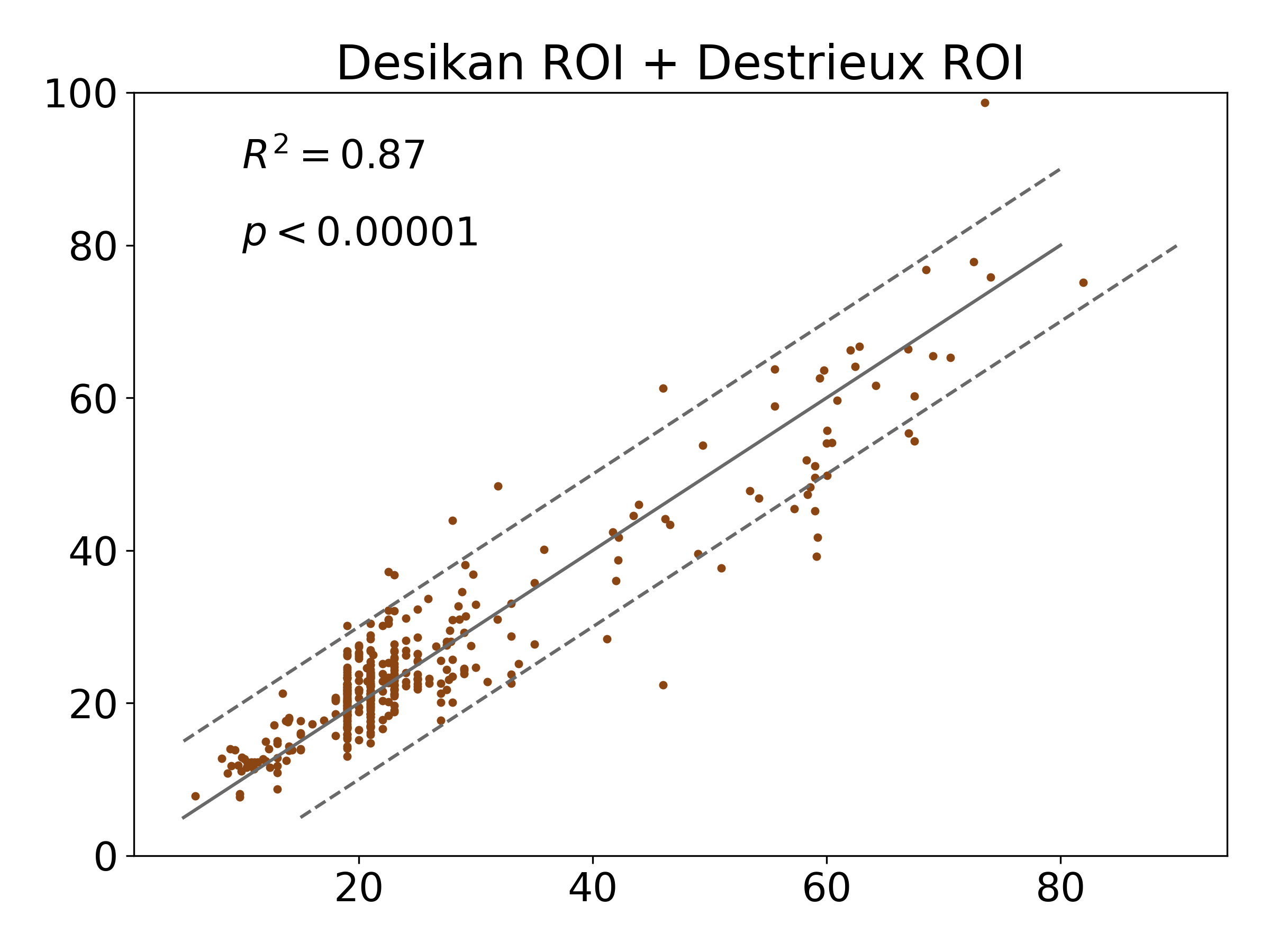}
        \label{fig:prob1_4_18}
    \end{minipage}\par\vspace{-1.3\baselineskip}
    \begin{minipage}{0.25\textwidth}
        \centering
        \includegraphics[width=0.98\linewidth]{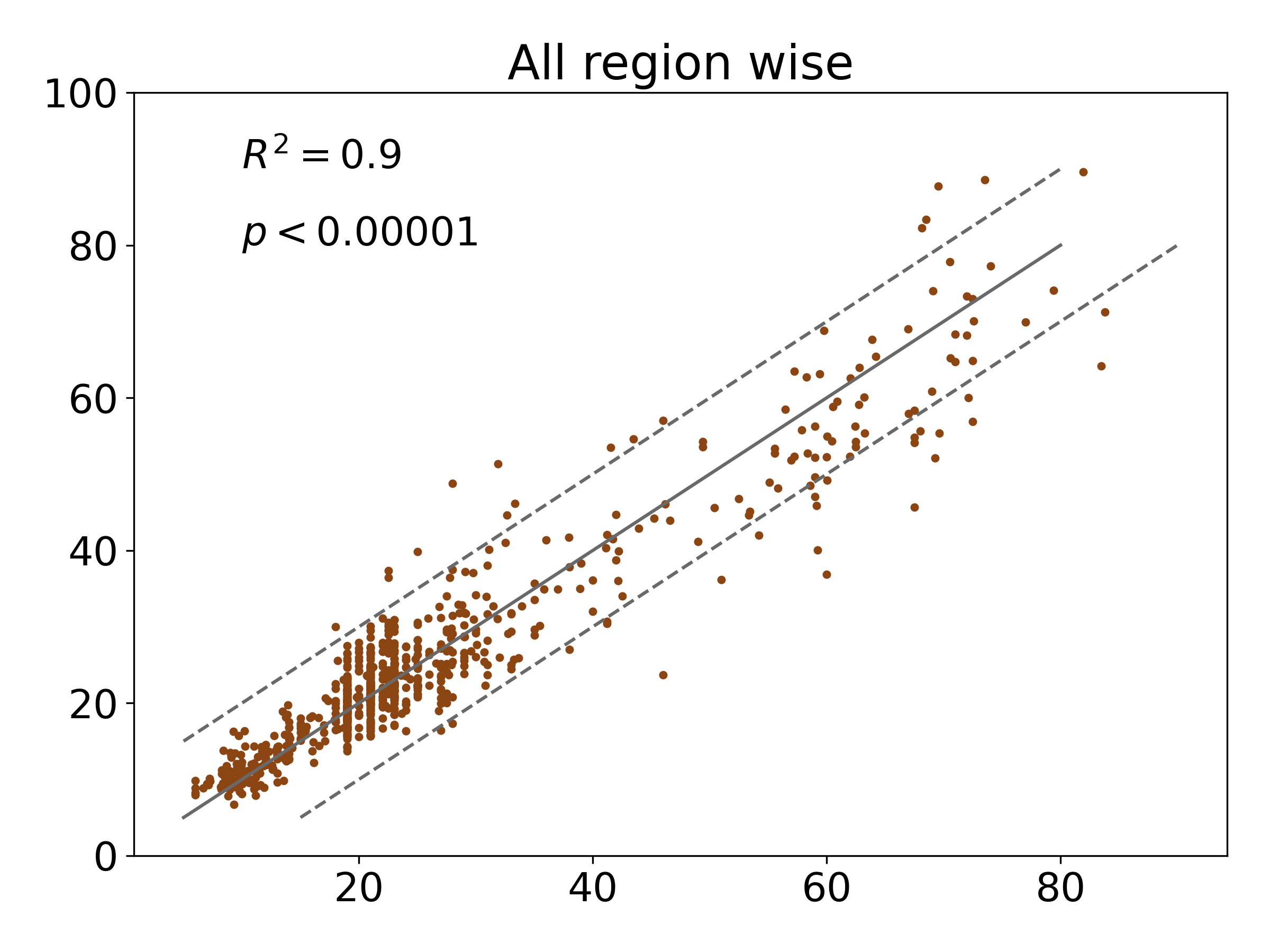}
        \label{fig:prob1_4_19}
    \end{minipage}%
    \begin{minipage}{0.25\textwidth}
        \centering
        \includegraphics[width=0.98\linewidth]{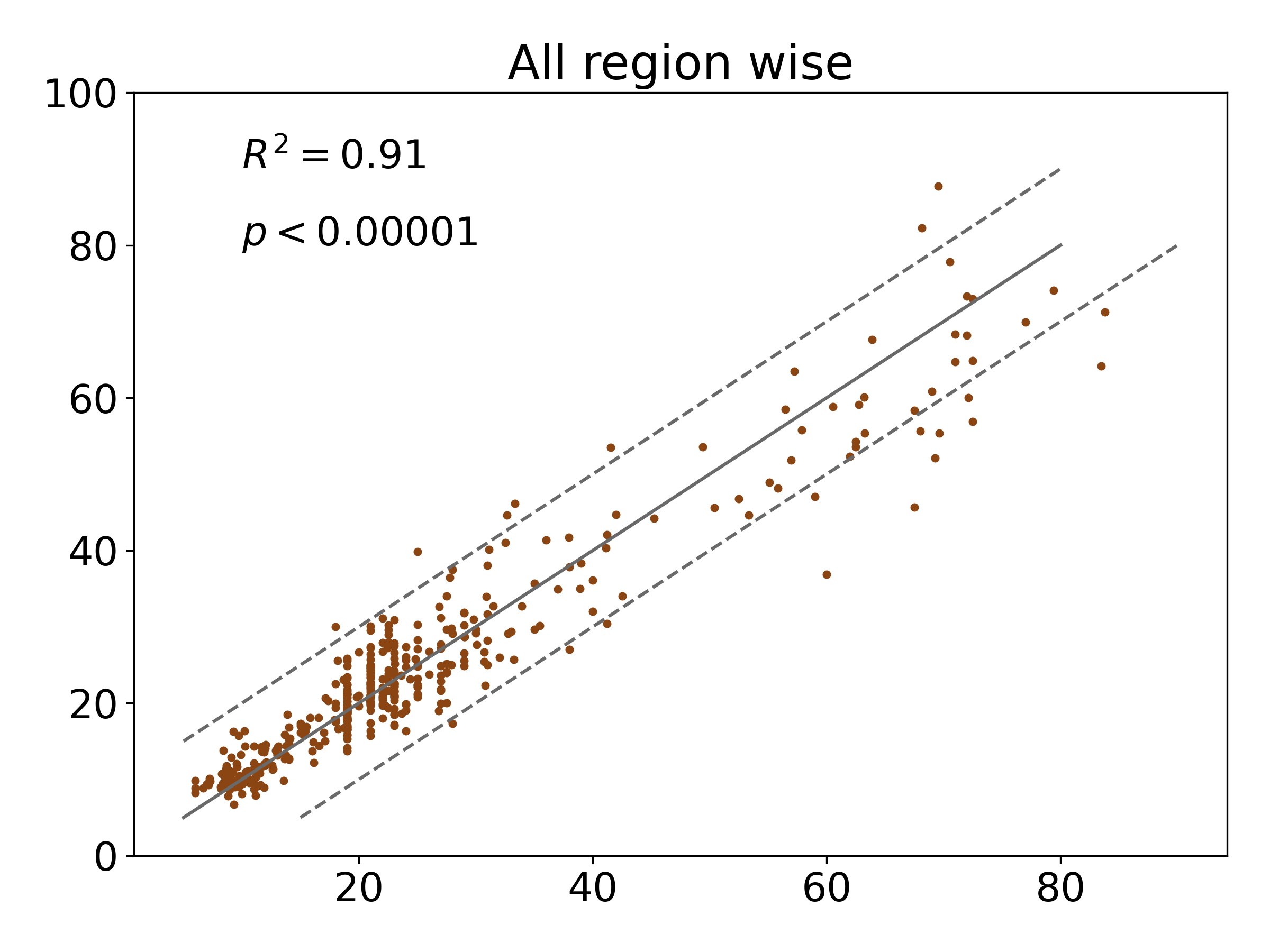}
        \label{fig:prob1_4_20}
    \end{minipage}%
    \begin{minipage}{0.25\textwidth}
        \centering
        \includegraphics[width=0.98\linewidth]{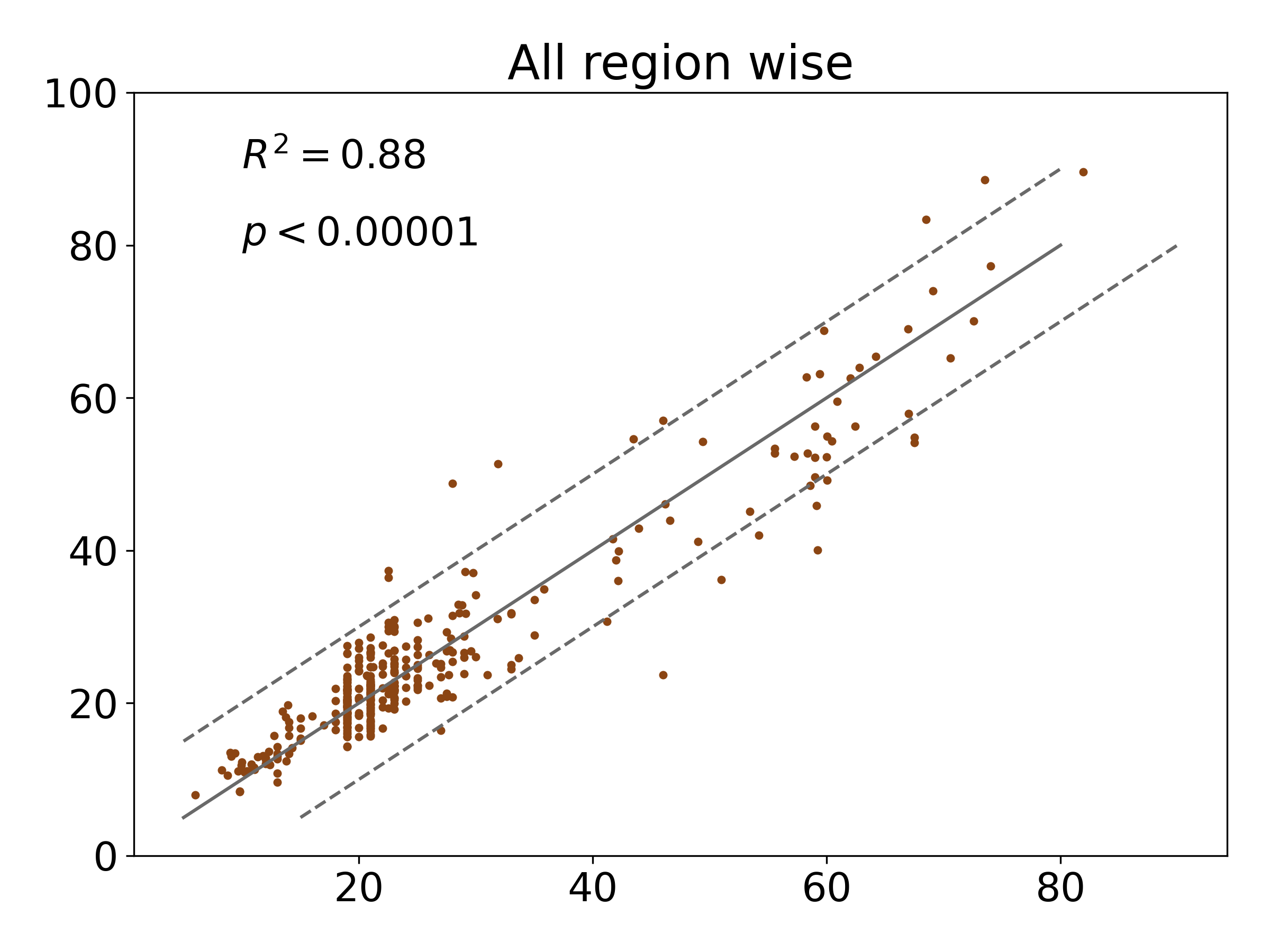}
        \label{fig:prob1_4_21}
    \end{minipage}\par\vspace{-1.3\baselineskip}
    \caption{(from left to right) Scatter plots showing the chronological age (years) vs. estimated brain age (years) of all healthy test subjects, the male healthy test subjects, and the female healthy test subjects using different MRI-derived region-wise features. }
        \label{fig_scatter_plots}  
\end{figure}

\begin{figure}[h!]
     \begin{minipage}{0.61\textwidth}
        \centering
        \includegraphics[width=0.99\linewidth]{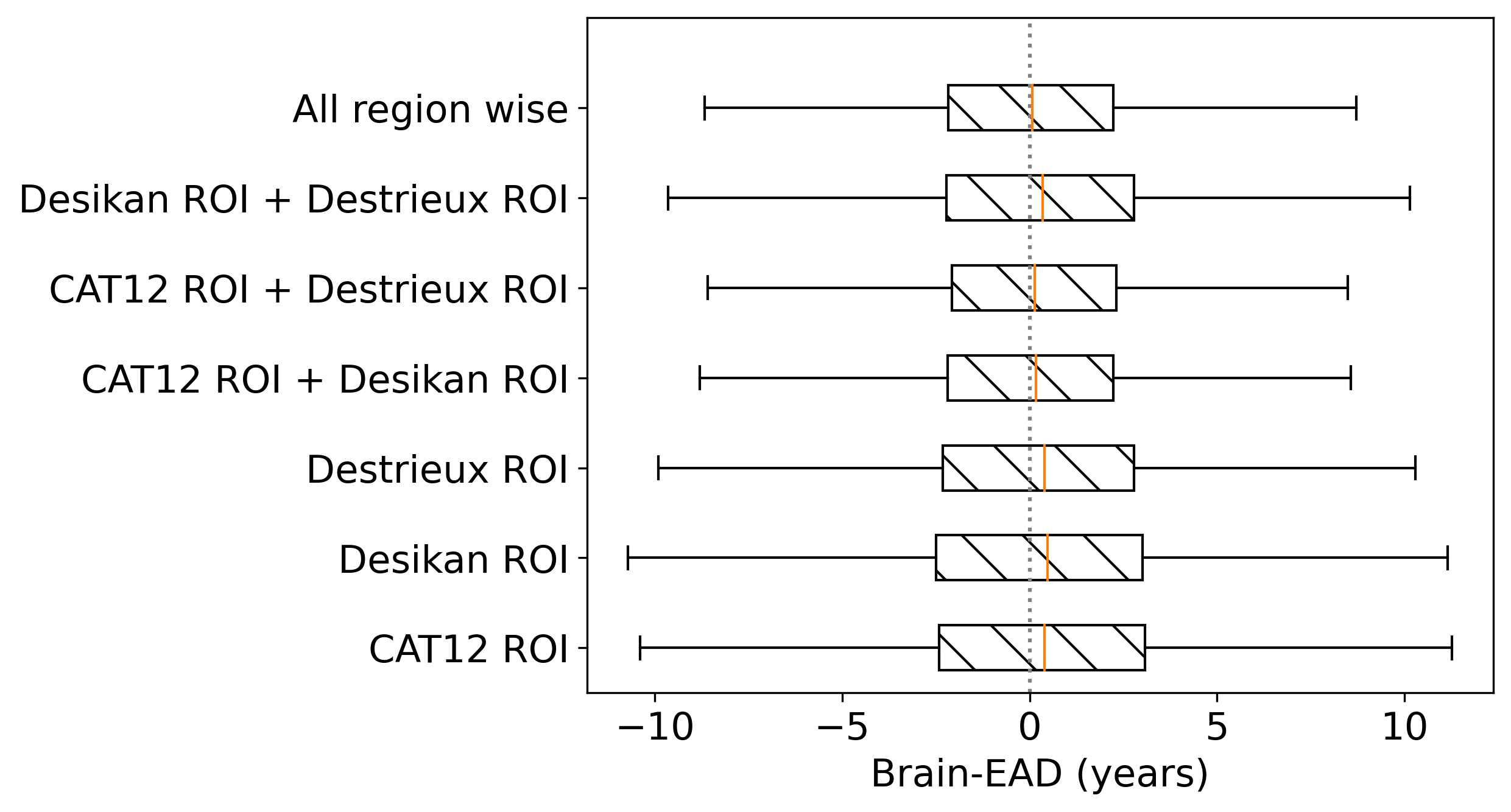}
        \label{fig:prob1_5_1}
    \end{minipage}%
    \begin{minipage}{0.39\textwidth}
        \centering
        \includegraphics[width=0.98\linewidth]{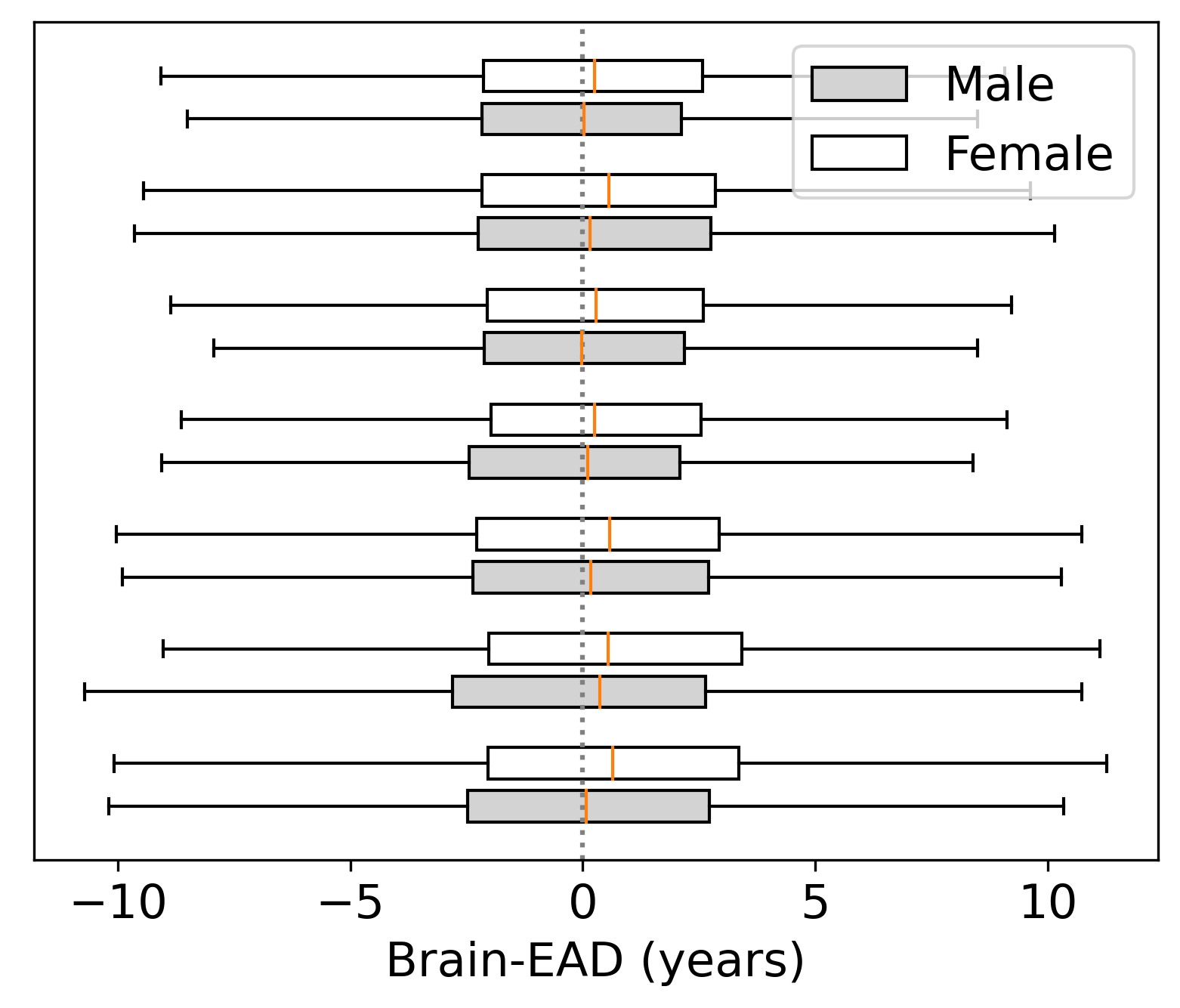}
        \label{fig:prob1_5_2}
    \end{minipage}
    \caption{(from left to right) Box plots showing the brain-EAD using the region-wise features of the independent, healthy test set and the male and female hold-out test sets.}
        \label{fig_box_plots}  
\end{figure}

We present scatter plots to show the difference between the chronological age and the estimated brain age using region-wise MRI-derived features in Figure~\ref{fig_scatter_plots}. This relationship is shown for both the male (number of samples $= 434$) and female (number of samples $= 359$) test participants. 
Box plots in Figure~\ref{fig_box_plots} show the brain-EAD using different region-wise features for both genders.
We test our BAE model on the MRI-derived features of healthy individuals, as shown in the right sub-table in Table~\ref{tbl_regression2}.

\begin{table}[h!]
    \centering
    \footnotesize
    \resizebox{0.85\textwidth}{!}{
    \begin{tabular}{lp{1cm}p{1cm}p{1cm}p{1cm}p{1cm}p{0.8cm}|p{1cm}p{1cm}p{0.6cm}}
    \toprule
        &\multicolumn{3}{c}{Male}&\multicolumn{3}{c}{Female} & \multicolumn{3}{c}{Complete Data}\\ 		
		\cmidrule(lr){2-4} \cmidrule(lr){5-7} \cmidrule(lr){8-10}
        \multirow{1}{2cm}{MRI features} & \multirow{1}{1.2cm}{MAE $\downarrow$} & \multirow{1}{1.3cm}{RMSE $\downarrow$} & \multirow{1}{0.8cm}{$R^2$ $\uparrow$} & \multirow{1}{1.2cm}{MAE $\downarrow$} & \multirow{1}{1.3cm}{RMSE $\downarrow$} & \multirow{1}{1cm}{$R^2$ $\uparrow$} & \multirow{1}{1.2cm}{MAE $\downarrow$} & \multirow{1}{1.3cm}{RMSE $\downarrow$} & \multirow{1}{0.8cm}{$R^2$ $\uparrow$} \\
        \midrule \midrule
        \multirow{1}{*}{CAT12 ROI}
         &3.9 & 5.68 & 0.86 & 4.17 & 6.11 & 0.81 & 3.94 & 5.32 & 0.87 	\\
          \cmidrule{2-7} \cmidrule{8-10}
        \multirow{1}{*}{Desikan ROI}
        &4.39 & 6.93 & 0.79 & 4.03 & 5.69 & 0.83  & 4.23 & 6.4 & 0.81 	\\
          \cmidrule{2-7} \cmidrule{8-10}
        \multirow{1}{*}{Destrieux ROI}
        &3.98 & 6.05 & 0.84 & 3.81 & 5.51 & 0.84  & 3.9 & 5.81 & 0.84	\\
          \cmidrule{2-7} \cmidrule{8-10}
        \multirow{1}{*}{CAT12 ROI + Desikan ROI}
        &3.29 & 4.94 & 0.89 & 3.54 & 5.11 & 0.87   & 3.4 & 5.02 & 0.88 	\\
          \cmidrule{2-7} \cmidrule{8-10}
        \multirow{1}{*}{CAT12 ROI + Destrieux ROI}
        &3.24 & 4.78 & 0.90 & 3.44 & 4.97 & 0.87   & 3.33 & 4.87 & 0.89	\\
         \cmidrule{2-7} \cmidrule{8-10}
        \multirow{1}{*}{Desikan ROI + Destrieux ROI}
        &3.89 & 5.92 & 0.85 & 3.63 & 5.14 & 0.87  & 3.77 & 5.58  & 0.86 	\\
         \cmidrule{2-7} \cmidrule{8-10}
        \multirow{1}{*}{\textbf{All region}}
        &\textbf{3.19} & \textbf{4.67} & \textbf{0.91} & \textbf{3.32} & \textbf{4.79} & \textbf{0.88}   & \textbf{3.25} & \textbf{4.73} & \textbf{0.90}	\\
        \bottomrule
         \end{tabular}
    }\medskip
    \caption{Performance comparison of the brain age estimation framework for different MRI-derived region features of the male and female healthy test set (left sub table) and complete  (containing male and female combined) healthy test set (right sub table).}
    \label{tbl_regression2}
\end{table}

Integrating all region-wise feature metrics from T1-w MRI improved the BAE framework, resulting in lower MAE and RMSE than using individual feature metrics. The highest performance with individual feature metrics was achieved using the Destrieux-ROI features because we utilized seven cortical measurements or explanatory variables for $148$ brain regions, unlike previous studies that used fewer variables~\cite{BEHESHTI_BA_2022,LIU_2022_BA}. Conversely, the individual global volumetric measurements, CAT12 ROI, of the segmented GM and CSF achieved lower performance (higher MAE). The improved performance of our BAE model is also because of an overall greater sample size ($m = 3965$) compared to the previously known models \cite{Cole2017_BA_CNN,LIU_2022_BA,Jonsson_2019_BA} and wider age range compared to the state-of-the-art BAE models \cite{Aycheh_2018_bae_region,Baecker_2021_bae_reg,BEHESHTI_BA_2022}.  
\par We also observed that the choice of machine learning model greatly affects the performance of BAE using different region-wise MRI-derived features. Table~\ref{tbl_regression} shows that Generalized Linear Model (GLM) performed better on Desikan ROI and Destrieux ROI region-wise features compared to Linear Regression (LR), linear Support Vector Regression (SVR), and Relevance Vector Regression (RVR). On the contrary, RVR improved accuracy on the CAT12 ROI features compared to LR, SVR, and GLM. These results are consistent with the BAE models in~\cite{LEE_2021_BA,mishra_2021_BA_review}. 
\par The brain-EAD of the healthy test set is evident from Figure~\ref{fig_box_plots} using different region-wise features. It shows that integrating the three region-wise structural measurements decreases brain-EAD ($\mu \approx 0$). While, individually, CAT12 ROI and Desikan ROI showed similar results with the highest brain-EAD. Similarly, the scatter plots between the chronological age and brain age in Figure~\ref{fig_scatter_plots} show that the proposed model generalizes well for adolescents and adults (lesser Brain-EAD), while outliers exist among older healthy subjects because of the non-uniform age distribution across the different age groups. Another important observation is that the BAE model overestimates for younger subjects while underestimates for older subjects~\cite{TAYLOR_2022_BAE,Cole2017_BA_CNN} as shown in Figure~\ref{fig_scatter_plots}.  
\par We tested the proposed model separately on male and female subjects and found that the model performed slightly better for male participants ({\textsc mae} = $3.19$ years) compared to the female subjects ({\textsc mae} = $3.32$ years). 
This difference in performance is also observed in~\cite{franke_BA_2019} because of both genders' local and global brain anatomy and the normal aging trajectory~\cite{Farokhian_2017_aging,Sanford_bae_2022}. Finally, in the case of CAT12 ROI, the GM volume shows a strong -ve correlation with age ($r \approx -0.5$), while CSF volume shows a strong +ve correlation with age ($r \approx 0.5$). These observations are consistent with previous studies indicating that GM volume decreases with age, while CSF volume increases gradually with age~\cite{Farokhian_2017_aging,Hafkemeijer_2014}.
\par The present study has three limitations. First, besides using an extensive collection of healthy brain MRIs aged 6-86 years, the participants' age distribution is right-skewed compared to \cite{LIU_2022_BA,Basodi2022_BA}. A future research direction can include more adult and elder healthy training samples to explore the impact on the accuracy of the BAE model. Second, we only used the global volumetric features on the Neuromorphometrics atlas and did not explore other voxel-based atlases like Suit and Cobra. Third, the current approach only uses T1-w brain MRIs for brain age estimation and does not consider multimodal data such as fMRI used recently in \cite{Basodi2022_BA,TAYLOR_2022_BAE}.  

\vspace{-0.3cm}
\section{Conclusion}\label{sec:conclusion}
In this paper, we integrated the different region-wise features, i.e., global volumetric and cortical measurements of different parcellations, derived using T1-w healthy brain MRIs to develop a BAE framework using the novel benchmark dataset, OpenBHB. We evaluated the performance of our model and demonstrated that fusion of the different region-wise metrics results in improved BAE accuracy ($\textsc{mae} = 3.25$ years). These features, on their own, achieved comparatively lesser BAE performance ($\textsc{mae} = 4.23$ years) depending on the number of anatomical brain regions of interest and the different volumetric measurements. Our results demonstrate that robust BAE frameworks can be constructed by integrating the three different region-wise metrics (CAT12 ROI, FreeSurfer Desikan ROI, and FreeSurfer Destrieux ROI) derived from T1-w MRIs of healthy controls.

\bibliographystyle{splncs04}
\bibliography{references}

\end{document}